\documentclass[journal]{IEEEtran}
\usepackage{amsbsy,amsmath,amssymb,epsfig,bbm,mathrsfs,multirow,amsthm,subcaption,caption,graphicx,epstopdf,multicol,lipsum,float}
\usepackage{algorithm}
\usepackage{algorithmic}
\usepackage{tikz}
\usetikzlibrary{arrows,calc}

\tikzstyle{line} = [draw, -latex']

\begin{document}

\title{Shipper Cooperation in Stochastic Drone Delivery: A Dynamic Bayesian Game Approach}

\author{Suttinee~Sawadsitnag, Dusit~Niyato,~\IEEEmembership{Fellow,~IEEE,},~Puay-Siew~Tan, Ping~Wang,~\IEEEmembership{Senior~Member,~IEEE}, and Sarana~Nutanong,~\IEEEmembership{Member,~IEEE} 

\thanks{S. Sawadsitang and D. Niyato are with the School of Computer Science and Engineering, Nanyang Technological University, Singapore, 639798, e-mail: (e-mail: suttinee002@e.ntu.edu.sg; dniyato@ntu.edu.sg; wangping@ntu.edu.sg).}
\thanks{P.-S. Tan is with the Singapore Institute of Manufacturing Technology, Singapore 638075 (e-mail: pstan@simtech.a-star.edu.sg).}
\thanks{P. Wang is with the Department of Electrical Engineering and Computer Science, York University, Canada (e-mail: pingw@yorku.ca).}
\thanks{S. Nutanong is with the Vidyasirimedhi Institute of Science and Technology,
Rayong 21210, Thailand (e-mail:,snutanon@vistec.ac.th).}
\thanks{Color versions of one or more of the figures in this paper are available online at http://ieeexplore.ieee.org.}}

\markboth{Journal of \LaTeX\ Class Files,~Vol.~14, No.~8, August~2015}%
{Shell \MakeLowercase{\textit{et al.}}: Bare Demo of IEEEtran.cls for IEEE Journals}
\maketitle

\begin{abstract}
With the recent technological innovation, unmanned aerial vehicles, known as drones, have found numerous applications including package and parcel delivery for shippers. Drone delivery offers benefits over conventional ground-based vehicle delivery in terms of faster speed, lower cost, more environment-friendly, and less manpower needed. However, most of existing studies on drone delivery planning and scheduling focus on a single shipper and ignore uncertainty factors. As such, in this paper, we consider a scenario that multiple shippers can cooperate to minimize their drone delivery cost. We propose the Bayesian Shipper Cooperation in Stochastic Drone Delivery (BCoSDD) framework. The framework is composed of three functions, i.e., package assignment, shipper cooperation formation and cost management. The uncertainties of drone breakdown and misbehavior of cooperative shippers are taken into account by using multistage stochastic programming optimization and dynamic Bayesian coalition formation game. We conduct  extensive performance evaluation of the BCoSDD framework by using  customer locations from Solomon benchmark suite and a real Singapore logistics industry. As a result, the framework can help the shippers plan and schedule their drone delivery effectively.

\end{abstract}


\section{Introduction}

The market size of e-commerce has rapidly increased the demand of logistics industry. The global e-commerce logistics market will expand from US\$122.2 billion in 2014 to US\$781 billion in 2024~\cite{ref_market}. Accordingly, a package delivery service becomes the most important business function in logistics. Suppliers, which can be referred to as shippers, need to handle massive package delivery demands while maintaining high customer satisfaction. While majority of the package delivery is done through ground-based vehicles, e.g., trucks, ariel delivery using drones emerges as a promising solution because of the cost efficiency, reliability, and energy consumption. Drones have been on trial and adopted for small package delivery by many companies such as Amazon (2013), DHL (2013), Alibaba (2015), JD.com (2017), and Japan Post (2018). However, drone delivery is not always an optimal solution. It has some major limitations including limited flying distance, limited time, and weight capacity. Therefore, drone delivery is expected to be combined with other delivery services such as outsourcing package delivery to a third-party carrier. This combined delivery becomes an appropriate option for shippers to meet customers' needs while maintaining low cost and high reliability. 


Alternatively, logistics business partnership becomes a viable choice of reducing cost and improving delivery efficiency. Multiple shippers can cooperate and form a pool of delivery resources including depots, manpower, and drones~\cite{lee2014}. For example, a cooperative shipper can use vehicles to serve its customers and when the vehicles are not used, they can be utilized by the other cooperative shippers. This opportunistic logistics resource sharing can clearly improve the resource utilization and hence profitability. 

With the cooperation among shippers with drones, the shippers need to answer the following questions: (i) should the shippers cooperate with other shippers? (ii) which shippers should cooperate with? (iii) how can the shippers with cooperation absorb the delivery cost in a fair manner? and (iv) how many drones are required and how to assign packages to the drones or to use an outsource carrier. To address these questions, we therefore propose the Bayesian Shipper Cooperation in Stochastic Drone Delivery (BCoSDD) framework for helping shippers to strategize their delivery planning and resource sharing~\cite{ref_vtc_fall_2018}. The contributions of this paper are summarized as follows. 

\begin{itemize}
\item In the proposed BCoSDD framework, we introduce the package assignment based on multistage stochastic programming optimization. The package assignment takes the uncertainty of the drone breakdown into account to assign packages to be delivered by drones with the lowest cost while meeting all customers' demand. While the original problem of the package assignment is non-linear, we reformulate the problem into a linear one that can be solved more efficiently.
\item We propose a shipper cooperation management. Specifically, a cooperation decision making process of the shippers is modeled as a static and a dynamic Bayesian coalition formation game, respectively. The static Bayesian coalition formation game is applied when the belief about the shippers is known a priori. Alternatively, the dynamic Bayesian coalition formation game is applied when the shippers have to observe and learn behaviors of the other shippers without complete belief. We analyze the stability of the shipper cooperation analytically and also present an algorithm to reach the stable solution. Moreover, in reality, some shippers may misbehave in the cooperation, e.g., a shipper may not deliver a package as it is assigned, the algorithm is thus designed to learn the shippers' behavior and allow honest shippers to form the cooperation effectively.
\item Among the shippers cooperating with each other, the incurred cost will be distributed. We therefore propose  cost management to allow the cooperative shippers to share the cost of package delivery. The Shapley value technique is adopted to reach a fair cost sharing.
\end{itemize}
In summary, the proposed BCoSDD framework is useful for the shippers not only for their logistics resource and delivery planning, but also for reaching an optimal business partnership through cooperation to share the resources efficiently. The ultimate goals are to achieve the minimum cost and stable cooperation strategy by the shippers.

This paper is organized as follows. Section~\ref{sec_related} presents the literature review of the vehicle routing problem, drone package delivery, and the cooperation in package delivery. Section~\ref{sec_system} describes the system model, which consists of three components, i.e., package assignment, shipper cooperation, and cost management. The package assignment and the shipper cooperation are discussed in detail in Section~\ref{sec_package} and Section~\ref{sec_static_game}, respectively. After that, the dynamic algorithm for shipper cooperation formation is presented in Section~\ref{sec_dynamic_game}. Section~\ref{sec_experiment} explains the experiment setting and gives performance evaluation results of the proposed framework. Section~\ref{sec_conclusion} concludes the paper. 

\section{Related Work}
\label{sec_related}

Since goods distribution is one of the major tasks in supply chain, a number of studies have been conducted to address the package delivery and vehicle routing problem (VRP)~\cite{ref_greensurvey}. However, most of existing studies concentrate on ground-based vehicle delivery, i.e., by motorcycle, car, and truck, as they are convenient, efficient, and cost-effective. In the literature, there are VRP surveys for ground-based vehicle delivery. Each of the surveys considers a specific aspect of the problem, for example, VRP with time windows~\cite{ref_survey_1988}, pick-up and delivery VRP~\cite{ref_survey_2007-pickup}, dynamic VRP~\cite{ref_survey_2013-dynamic} and stochastic VRP~\cite{ref_survey_1996}, and solution algorithms~\cite{ref_survey_1992}, \cite{ref_survey_2000}.

Until recently, an unmanned aerial vehicle (UAV), as known as a drone, has emerged as an alternative package delivery mode. This is due to the fact that the reliability has been improved while the cost has been reduced substantially. As such, the VRP is being considered again for the drone delivery. Sundar {\em et~al.}~\cite{UAV-refuel} proposed an approximation algorithm for a single drone routing. The drone is assumed to depart from a depot and visit multiple locations before returning to the origin. The drone is allowed to stop over at different locations to refuel or recharge its energy. However, this work does not address the package delivery directly, and the proposed scheme is suitable only for surveillance applications. Murray and Chu~\cite{sidekick} studied a package delivery problem by using the combination of one drone and one truck. Since the drone may not deliver a package to every customer due to limited payload weight and/or flying distance, the truck can be used to serve these customers. Optimization problems are proposed to minimize the delivery time. Ferrandez {\em et~al.}~\cite{k_mean} also considered the joint truck and drone delivery problem. They proposed mathematical formulations for closed-form estimations. The K-mean algorithm and genetic algorithm are used to obtain the delivery solution. 

Dorling {\em et~al.}~\cite{drone_delivery} considered only drones for package delivery planning. They proposed two drone routing optimization problems given the constraints of capacity, battery weight, changing and payload weight. The objectives are (i) to minimize the cost subject to delivery time constraint and (ii) to minimize the time subject to the budget cost constraint. Then, they proposed the simulated annealing (SA) heuristic algorithm to address the multi-trip drone routing problem. Hong {\em et~al.}~\cite{drone_obstacle} also studied the drone-only package delivery problem. Unlike other previous studies that assume drones to fly only over roads and street, the problem focuses on the flying direction of drones, which can avoid barriers and obstacles automatically. The problem is formulated as mix-integer programming, and a heuristic algorithm is used to find the solution. However, all the above studies do not consider the uncertainty of a drone, e.g., drone breakdown, which is common in practice. In our previous work~\cite{maggie_TITS}, a joint truck and drone delivery problem was formulated as a three-stage stochastic programming problem with the objective to minimize a set of costs while meeting the customers' delivery demand.

Since the cost is a sensitive factor in logistics industry, partnership and cooperation among shippers are seen as a viable solution to improve resource utilization in package delivery problem. Krajewska {\em et~al.}~\cite{ref_game1} introduced the joint routing and cooperation among shippers for ground-based package delivery. They used the Shapley value to fairly distribute cost among shippers. Later on, the cooperative game theory was used in many ground-based package delivery studies~\cite{ref_game2},~\cite{ref_game3},~and \cite{ref_game4}. However, all of them concentrate on presenting different methods for cost sharing. The review of cost sharing for shipper cooperation in package delivery was presented by Guajardo and Ronnqvist~\cite{ref_game_survey}. Recently, Kaewpuang {\em et~al.}~\cite{ref_pboo} proposed a framework for shipper cooperation for the capacitated vehicle routing problem with time window. Instead of focusing on the cost sharing methods, they considered the overlapping coalition formation among shippers, i.e., one shipper may join multiple resource pools at the same time. The Myerson value was adopted as the solution to distribute cost among cooperative shippers. 

However, it is essential to note that all previous studies of the shipper cooperation in package delivery are dedicated to the ground-based delivery. Moreover, none of them considers the shippers' misbehavior. The literature also misses the study of jointly optimizing package assignment, shipper cooperation, and cost management. Therefore, they are the focus of this paper in which the uncertainty in drone breakdown is also taken into consideration.





\section{System Model and Assumptions}
\label{sec_system}

This section presents the proposed BCoSDD framework in detail. The framework considers multiple shippers, and each shipper is rational and self-interested. All the shippers have the same objective that is to minimize their delivery cost while the packet delivery must be done in a specific time slot, e.g. all packages are delivered by the end of the day. The shippers can use drones to deliver a package to a customer or outsource a package delivery to a third-party carrier. The drone flies from a depot to serve the customer and returns to the depot. If a shipper outsources the package delivery to a carrier, the shipper pays a fixed cost, i.e., a fee, to the carrier. When a shipper or a coalition of  shippers deliver multiple packages, the delivery cost is assumed to be cheaper than that of outsourcing package delivery. Nonetheless, outsourcing package delivery to a carrier is still needed to meet customers' demand in some cases, for example, when drones cannot serve customers which are not in the flying coverage area. 

Furthermore, the shippers can cooperate and create a resource pool to minimize their delivery cost. Once the resource pool is created, a shipper can let other cooperative shippers the same pool deliver a package. In this case, the package can be transferred from one depot to another depot of different cooperative shippers to facilitate the use of drone for delivery, i.e., drone sharing. With the resource pool, the incurred cost will be distributed by the cooperative shippers forming the same pool in a fair manner.

\begin{figure}[b]

\framebox[0.5\textwidth]{\centering
\begin{tikzpicture}[->,>=stealth',shorten >=0.5pt,auto,node distance=0.5cm,
 main node/.style={rectangle,fill=blue!6,draw,
 font=\scriptsize,minimum size=3mm}]
 \node[main node] (D1) at (0,0){\begin{tabular}{c}\textbf{Shipper Cooperation}\\ \framebox[1.1\width]{Merge \& Split Algorithm} \end{tabular}};
 \node[main node] (D2) at (-2.1,-1.3) {\begin{tabular}{c} \textbf{Package Assignment} \\ \framebox[1.1\width]{Optimization (MIP)} \end{tabular} };
 \node[main node] (D3) at (2.1,-1.3) {\begin{tabular}{c}\textbf{ Cost Management}\\\framebox[1.1\width]{Shapley Value} \end{tabular} };
\draw[->, to path={-| (\tikztotarget)}] (D1) edge (D2) ;
 \path [line] (D2) |- (D3);
 \path [line] (D3) |- (D1); 
 \node at (2.8,-0.5){\tiny\begin{tabular}{c}Cost \\ Sharing \end{tabular}} 
;
	\node at (0.3,-1.6){\tiny \begin{tabular}{c} Optimal \\ Solution \vspace{-1em}\end{tabular}}; 
	\node at (-2.8,-0.5){\tiny \begin{tabular}{c} Coalitional\\ Structure \end{tabular}}; 
\end{tikzpicture} 
}
\caption{The shipper cooperation in static BCoSDD framework.}
\label{f_comp}
\vspace{-1.5em}
\end{figure}
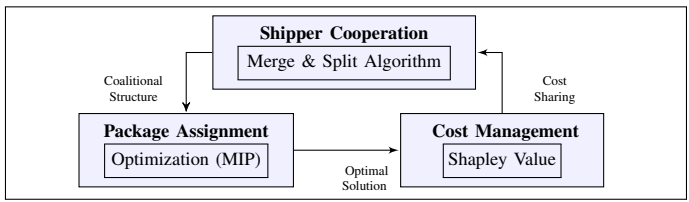

The BCoSDD framework is composed of three components as shown in Figure~\ref{f_comp}. We first address the package assignment component for an individual shipper or a coalition of cooperative shippers, also called a coalition of shippers, by an optimization technique. Next, we present the cost management component which is done after the solution of the package assignment problem is obtained. After that, the shipper cooperation component is used to decide how the shippers should cooperate and form a coalition and a resource pool. The interactions among these components are as follows.
\begin{enumerate}
	\item First, given a set of coalitions of cooperative shippers, i.e., a coalition structure, the package assignment is performed to assign packages to drones or to outsource packages to a carrier. 
	\item Second, the optimal solution of the package assignment is then used by the cost management to divide the package delivery cost among the cooperative shippers in the same group.
	\item Third, given the cost divided among the cooperative shippers, the shipper cooperation is performed, i.e., through the merge-and-split algorithm, to achieve the stable coalition structure.
\end{enumerate}


\section{Package Assignment}
\label{sec_package}

In this section, the package assignment is discussed in the detail. We first introduce how the package assignment process works. Then, we present the formulations of the package assignment optimization. 

\subsection{Package Delivery Operation and Scenario Tree}

Let $\mathcal{P}= \{p_1,p_2,\dots,p_{|\mathcal{P}|} \}$ be a set of all shippers in the system, where $|\mathcal{P}|$ denotes the total number of shippers. Each shipper $p$ has one depot, a set of customers denoted by $C_p$, and a set of drones denoted by $D_p$. Without loss of generality, each customer has one package to be delivered. Note that if a customer has two or more packages to be delivered, we can consider that there are multiple customers located at the same location. 
 
The package assignment is executed for one given coalition of cooperative shippers. Let $\mathcal{S}$ be a set of shippers that cooperate with each other, creating a resource pool and forming a coalition $\mathcal{S}$, i.e. $\mathcal{S}\subset \mathcal{P}$. The shippers in the same coalition allow their drones to deliver packages of each other. In this case, a shipper can transfer its packages from its depot to the depots of  other shippers. Accordingly, the packages of the customers of the shippers in the same coalition will be delivered by any of the drones. For example, for $\mathcal{S}=\{p_1,p_2,p_3\}$ i.e., shippers $p_1$, $p_2$, and $p_3$ cooperate and form a coalition, the set of customers for this coalition is $\mathcal{C} = C_1\cup C_2\cup C_3$ and the set of drones in this coalition is $\mathcal{D} = D_1 \cup D_2 \cup D_3$. In other words, $\mathcal{C} = \{c_1,c_2,\dots,c_{|\mathcal{C}|}\}$ and $\mathcal{D} = \{d_1,d_2,\dots,d_{|\mathcal{D}|}\}$ denote the sets of customers and drones in the pool, where $|\mathcal{C}|$ and $|\mathcal{D}|$ represent the total number of customers and the total number of drones, respectively. As the drones are shared, their utilization and efficiency can be increased. Note that if the shipper does not cooperate with any other shippers, the coalition contains only one shipper in which the same package assignment still applies. 

We assume that a drone can depart and return to the same depot only and cannot switch the depot. This is, for example, due to the flying program embedded in the drone. However, although the reliability of drones have been improved significantly, they are still subject to breakdown and failure, e.g., from bad weather conditions. A drone breakdown happens randomly. Thus, when a drone delivers a package to a customer, it can be successful or fail. If there is no breakdown, the drone will continue to deliver the next customer until all the assigned packages are delivered. However, if there is a breakdown, the drone cannot be used to deliver the rest of the assigned packages. In this case, the shippers have to pay penalty such as compensation to the customers of the rest of the assigned packages. Note that shippers can outsource the package delivery to a third-party carrier. However, the cost of outsourcing is typically higher than that of using their drones. Nonetheless, outsourcing can still be used, for example, if the chance of drown breakdown is high because of looming bad weather conditions.


\subsection{Package Assignment Formulations}

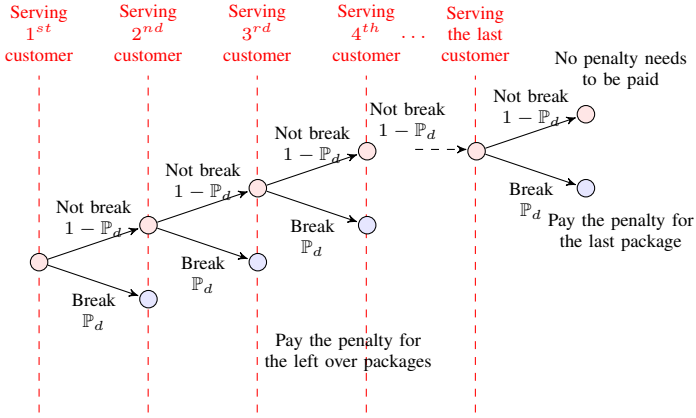
\begin{figure}
\hspace{-2em}
\scriptsize
\begin{tikzpicture}[->,>=stealth',shorten >=0.5pt,auto,node distance=1.7cm,
 red node/.style={circle,fill=red!10,draw, minimum size=0.5mm}, blue node/.style={circle,fill=blue!10,draw, minimum size=0.5mm}]

	\draw[-,red,dashed]  (0,-2) -- (0,2.5)node[above]{\begin{tabular}{c} Serving \\ $1^{st}$\\ customer \end{tabular} };
	\draw[-,red,dashed]  (1.45,-2) -- (1.45,2.5)node[above]{\begin{tabular}{c} Serving \\ $2^{nd}$\\ customer \end{tabular} };
	\draw[-,red,dashed]  (2.9,-2) -- (2.9,2.5)node[above]{\begin{tabular}{c} Serving \\ $3^{rd}$\\ customer \end{tabular} };
	\draw[-,red,dashed]  (4.35,-2) -- (4.35,2.5)node[above]{\begin{tabular}{c} Serving \\ $4^{th}$\\ customer \end{tabular} };
	\draw[-,red,dashed]  (5.8,-2) -- (5.8,2.5)node[above]{\begin{tabular}{c} Serving \\ the last\\ customer \end{tabular} };
	\draw[->,dashed]  (5,1.5) -- (5.7,1.5) node[above left] {\begin{tabular}{c} Not break \\ $1- \mathbb{P}_d$ \end{tabular} };
 \node[red node] (D1) {};
 \draw[-,red,dashed]  (5,3) node{$\dots$};
 
\node[red node] (D2) [ right of=D1, yshift=2em, xshift = -1em] {};
 \node[blue node] (D3) [ right of=D1, yshift=-2em, xshift = -1em] {};
\node[red node] (D4) [ right of=D2, yshift=2em, xshift = -1em] {};
 \node[blue node] (D5) [ right of=D2, yshift=-2em, xshift = -1em] {};
 \node[red node] (D6) [ right of=D4, yshift=2em, xshift = -1em] {};
 \node[blue node] (D7) [ right of=D4, yshift=-2em, xshift = -1em] {};
  \node[red node] (D8) [right of=D6, yshift=0em, xshift = -1em]{};
 \node[red node] (D9) [ right of=D8, yshift=2em, xshift = -1em] {};
 \node[blue node] (D10) [ right of=D8, yshift=-2em, xshift = -1em] {};
	\node[above right of= D9,yshift=-2.5em,xshift = -3em]{\begin{tabular}{c} No penalty needs \\ to be paid\end{tabular}};
	\node[below right of= D10,yshift=2.5em,xshift = -3em]{\begin{tabular}{c} Pay the penalty for \\ the last package\end{tabular}};
	\node[below right of= D5,yshift=0em,xshift = 0em]{\begin{tabular}{c} Pay the penalty for \\ the left over packages\end{tabular}};

 \path[every node/.style={inner sep=1pt}]
 (D1) edge node [anchor=center, above]{\begin{tabular}{c} Not break \\ $1- \mathbb{P}_d$ \end{tabular} } (D2) 
	 edge node [below=1mm] {\begin{tabular}{c} Break \\ $\mathbb{P}_d$ \end{tabular} } (D3)
(D2) edge node [below=1mm]{\begin{tabular}{c} Break \\ $\mathbb{P}_d$ \end{tabular} } (D5)
	edge node [anchor=center, above]{\begin{tabular}{c} Not break \\ $1- \mathbb{P}_d$ \end{tabular} } (D4)
	(D4) edge node [below=1mm]{\begin{tabular}{c} Break \\ $\mathbb{P}_d$ \end{tabular} } (D7)
	edge node [anchor=center, above]{\begin{tabular}{c} Not break \\ $1- \mathbb{P}_d$ \end{tabular} } (D6)
	(D8) edge node [below=1mm]{\begin{tabular}{c} Break \\ $\mathbb{P}_d$ \end{tabular} } (D10)
	edge node[anchor=center, above] {\begin{tabular}{c} Not break \\ $1- \mathbb{P}_d$ \end{tabular} } (D9);

\end{tikzpicture} 
\caption{An example of a scenario tree of a drone.}
\label{fig_sys_scenario}
\end{figure}

We formulate the package assignment problem as a multi-stage stochastic programming optimization. Figure~\ref{fig_sys_scenario} shows a scenario tree example of the stochastic model for one drone. In the first stage, the following activities happen, i.e., the drone is reserved for the delivery, selected packages are transferred to different depots, and the shippers outsource the package delivery to the carrier. From the second stage, each of the stages is associated with the delivery of one package of a customer sequentially in which the total number of stages is $|c|^{\text{th}}$. At the transition from the current stage to the next stage, there are two possible events, i.e., the delivery is successful or fails, which is represented by a branch. If the delivery is successful, the next stage ensues. However, if the delivery fails, the penalty is paid and the transition terminates as the drone cannot be used to deliver the rest of the assigned packages. Here, $\mathbb{P}_d$ denotes a breakdown probability of drone $d$, which can be calculated based on historical data.


The objective of the package assignment is to minimize the total delivery cost of a coalition of cooperative shippers, i.e., $\mathbb{O}^{ST}$. The multi-stage stochastic programming formulation is expressed as follows:

\noindent Minimize: 
\begin{align}
\mathbb{O}^{\mathrm{ST}}(\mathcal{S}) = \mathbb{O}^{\mathrm{DE}}+\sum_{d \in \mathcal{D}}\sum_{j=1}^{N_d} \left((1-\mathbb{P}_d)^{j-1} \mathbb{P}_dC^{(p)}(N_d-j+1)\right), \label{eq_sto_obj}
\end{align}
\noindent where
\begin{align}
\mathbb{O}^{\mathrm{DE}}=&\sum_{d\in \mathcal{D}} C^{(i)}_dW_{d} 
+ \hspace{-2em}\sum_{i \in \mathcal{C}, d\in \mathcal{D}, p \in \mathcal{S}}\hspace{-2em}\left( C^{(r)}_{p,i} + C^{(r)}_{i,p} \right) Y_{i,d,p} 
\nonumber\\
&+ \sum_{p \in \mathcal{S}}C^{(t)}_{p}T_p + \sum_{i \in \mathcal{C}}C^{(c)}_iZ_i, 
\label{eq_det_obj}
\end{align}
subject to (\ref{eq_det_initial}) - (\ref{eq_det_depot2}). Recall that the optimization formulation of the package assignment is for a given coalition $\mathcal{S}$. Nonetheless, we omit $\mathcal{S}$ in the expressions and notations for the rest of this section to simply the presentation. 

The delivery cost as expressed in (\ref{eq_sto_obj}) is composed of 
\begin{itemize}
\item the initial cost of drone $d$, i.e., $C^{(i)}_d$, 
\item the routing cost of drone $d$ to travel from the depot of shipper $p$ to customer $i$, i.e., $C^{(r)}_{p,i}$, 
\item the cost of shipper $p$ transferring packages among depots of the resource pool, i.e., $C^{(t)}_{p}$, 
\item the outsourcing cost per package, i.e., $C^{(c)}_i$ paid to a carrier, and 
\item the penalty cost when the drones fail to deliver a package, i.e., $C^{(p)}$.
\end{itemize}
There are seven decision variables in the formulation, including $W_{d}$, $Y_{i,d,p}$, $Z_i$, $T_p$, $M_{i,p,q}$, $B_{d,p}$, and $N_d$. The definitions of these decision variables are as follows. 
\begin{itemize}
\item $W_{d}$ is the indicator whether drone $d$ is used or not, i.e., if $W_{d} =1$, drone $d$ will be used in the delivery, and $W_{d}=0$ otherwise. 
\item $Y_{i,d,p}$ is the allocation variable. If $Y_{i,d,p} =1$, customer $i$ will be served by drone~$d$, and the drone will depart from depot~$p$, and $Y_{i,d,p} =0$ otherwise.
\item $Z_i$ is the indicator whether customer~$i$ will be served by a drone or not. If $Z_i =1$, customer $i$ will not be served by any drone, and $Z_i=0$ otherwise. 
\item $T_p$ is the indicator whether shipper~$p$ has to transfer packages from/to other shippers or not. If $T_p =1$, shipper $p$ will transfer packages to another cooperative shipper, and $T_p =0$ otherwise. 
\item $M_{i,p,q}$ is the indicator whether shipper $p$ transfers the package of customer~$i$ to shipper~$q$ or not. If $M_{i,p,q}=1$, shipper~$p$ transfers the package of customer~$i$ to shipper~$q$, and $M_{i,p,q}=0$ otherwise. 
\item $B_{d,p}$ is an auxiliary variable for imposing a drone to have only one departure and returning depot. 
\item $N_d$ is the total number of customers that will be served by drone $d$.
\end{itemize}

The package assignment is subject to the constraints in (\ref{eq_det_initial}) - (\ref{eq_det_depot2}). 
\begin{align}
&\sum_{i \in \mathcal{C}, p \in \mathcal{S} }Y_{i,d,p} \leq \Delta W_{d}, & \forall d \in \mathcal{D} \label{eq_det_initial}\\
&\sum_{i \in \mathcal{C},q \in \mathcal{S}}M_{i,p,q} \leq \Delta T_p, & \forall p \in \mathcal{S}\label{eq_det_tran1}\\
&\sum_{i \in \mathcal{C},q \in \mathcal{S}}M_{i,q,p} \leq \Delta T_p, & \forall p \in \mathcal{S}\label{eq_det_tran2}\\
&a_i \sum_{p \in \mathcal{S}}Y_{i,d,p} \leq f_d, & \forall i \in \mathcal{C}, d \in \mathcal{D} \label{eq_det_capacity}\\
&\sum_{d \in \mathcal{D}, p \in \mathcal{S} }Y_{i,d,p} + Z_i = 1, & \forall i \in \mathcal{C} \label{eq_det_allocation}
\end{align}
\begin{align}
&o_{i,p} - \sum_{ q \in \mathcal{S}}M_{i,p,q} \leq \Delta \sum_{d \in \mathcal{D}}\left(Y_{i,d,p} + Z_i \right),& \forall i \in \mathcal{C}, p \in \mathcal{S} \label{eq_det_belonging}
\end{align}

The constraint in (\ref{eq_det_initial}) ensures that the initial cost of a drone is paid when any packages are assigned to the drone. Note that $\Delta$ denotes a large number. The constraints in (\ref{eq_det_tran1}) and (\ref{eq_det_tran2}) ensure that the package transferring cost is paid when shippers transfer packages to the other shippers. The constraint in (\ref{eq_det_capacity}) ensures that the total weight of a package to be delivered by a drone does not exceed the capacity limit of the drone, where drone $d$ can carry a package with the weight up to $f_d$ kilograms. The constraints in (\ref{eq_det_allocation}) and (\ref{eq_det_belonging}) ensure that all the customer packages are assigned to a drone, or otherwise the outsourcing cost needs to be paid. We use variable $o_{i,p} = 1$ to define that customer $i$ belongs to shipper $p$, and $o_{i,p} = 0$ otherwise. We have the condition $\sum_{p \in \mathcal{P}}o_{i,p} = 1$ for all $i \in \mathcal{C}$. If $o_{i,p} = 1$, a drone that will deliver the package of customer $i$ must depart from the depot of shipper $p$. Otherwise, the package must either be transferred to a new depot or outsourced to a carrier. The constraints in (\ref{eq_det_limit_trip}) to (\ref{eq_det_limit_time}) ensure that the drone does not travel exceeding the coverage area, traveling distance limit, and traveling time limit, respectively. Here, $t_i$ is the serving time, i.e., the time that the drone spends to search and drop a package at the destination. The drone has flying distance limits, which are up to $e_d$ kilometers per trip and $l_d$ kilometers per day. The average flying speed of drone $d$ is denoted as $s_d$. Let $k_{i,j}$ denote the flying distance from location $i$ to location $j$. 

\begin{align}
&\sum_{p \in \mathcal{S}}Y_{i,d,p}\left( k_{p,i} + k_{i,p} \right) \leq e_d, & \forall i \in \mathcal{C}, d \in \mathcal{D} \label{eq_det_limit_trip} \\
&\sum_{i \in \mathcal{C}, p \in \mathcal{S}}\hspace{-0.8em}Y_{i,d,p}\left( k_{p,i} + k_{i,p} \right) \leq l_d, & \forall d \in \mathcal{D} \label{eq_det_limit_day}\\
&\sum_{i \in \mathcal{C},p \in \mathcal{S}}\left(\frac{k_{i,p}+k_{i,p}}{s_d} + t_i\right)Y_{i,d,p} \leq h_d, & \forall d \in \mathcal{D} \label{eq_det_limit_time}
\end{align}

The constraints in (\ref{eq_det_tran3}) to (\ref{eq_det_tran5}) are the package transferring constraints. The constraint in (\ref{eq_det_tran3}) ensures that if a package is transferred to a new depot, the drone must depart from the new depot. The constraint in (\ref{eq_det_tran4}) ensures that a package can only be transferred once. The constraint in (\ref{eq_det_tran5}) ensures that no package is transferred to the original depot. The constraints in (\ref{eq_det_depot1}) and (\ref{eq_det_depot2}) ensure that a drone can only depart from and arrive to only one depot. 

\begin{align}
&M_{i,p,q} \leq \sum_{d \in \mathcal{D}} Y_{i,d,q}, & \forall i \in \mathcal{C}, \forall p \in \mathcal{S}, \forall q \in \mathcal{S} \label{eq_det_tran3}
\\
&\sum_{p\in \mathcal{S}, q \in \mathcal{S}}M_{i,p,q} = 1, & \forall i \in \mathcal{C} \label{eq_det_tran4}
\\
&M_{i,p,p} = 0, & \forall i \in \mathcal{C}, \forall p \in \mathcal{S}\label{eq_det_tran5}\\
&\sum_{i \in \mathcal{C}}Y_{i,d,p} \leq \Delta B_{d,p}, & \forall d \in \mathcal{D}, \forall p \in \mathcal{S} \label{eq_det_depot1}\\
&\sum_{p \in \mathcal{S}} B_{d,p} = 1, & \forall d \in \mathcal{D} \label{eq_det_depot2}
\\
&\sum_{i \in \mathcal{C}, p \in \mathcal{S}}Y_{i,d,p} \leq N_d, & \forall d \in \mathcal{D} \label{eq_sto_1}\\
&0 \leq N_d \leq |C|, & \forall d \in \mathcal{D}\label{eq_sto_2}
\end{align}

The constraints in (\ref{eq_sto_1}) and (\ref{eq_sto_2}) limit the number of serving customers. The constraint in (\ref{eq_sto_1}) calculates the total number of customers that are served by drone $d$, i.e., $N_d$. The constraint in (\ref{eq_sto_2}) ensures that $N_d$ is a positive integer and must be less than the total number of customers.

We can obtain the solution of the package assignment by solving the above multi-stage stochastic programming optimization. However, it is a quadratic programming problem as the objective function is in a quadratic form. Since quadratic programming problems can be more complex to solve than linear programming problems, next we transform the quadratic programming problem to the linear programming problem.

\subsection{Linear Programming Model}
We reformulate the quadratic programming problem in~(\ref{eq_sto_obj}) to the linear programming problem by introducing three auxiliary variables and three auxiliary constraints. The linear objective function is presented in (\ref{eq_sto_obj_lin}).

\noindent Minimize: 
\begin{align}
\mathbb{O}^{\mathrm{ST}} = \mathbb{O}^{\mathrm{DE}} + \sum_{i \in \mathcal{C}, d \in \mathcal{D}}A_{i,d},\label{eq_sto_obj_lin}
\end{align}
subject to (\ref{eq_det_initial}) - (\ref{eq_sto_2}) and (\ref{eq_sto_lin_1}) - (\ref{eq_sto_lin_4}). $\mathbb{O}^{DE}$ can be calculated from (\ref{eq_det_obj}). 

The definitions of the three auxiliary variables are as follows. 
\begin{itemize}
\item $A_{i,d}$ is an auxiliary positive variable. The value of $A_{i,d}$ will be a part of breakdown penalty, which is incurred by drone $d$.
\item $X_{i,d}$ is an auxiliary binary variable. $X_{i,d}=1$ when the order of $i$ is less than or equal to the total number of drone $d$ to be used in the delivery. Otherwise, $X_{i,d}=0$.
\item $V_{i,d}$ is an auxiliary positive variable. The value of $V_{i,d}$ will be a part of breakdown penalty, which is incurred by drone $d$. 
\end{itemize}

\begin{align}
&\sum_{i \in \mathcal{C}}X_{i,d} = N_d, & \forall d \in \mathcal{D} \label{eq_sto_lin_1}\\
&X_{i,d} \geq X_{j,d}, & \forall i,j \in C, i < j, \forall d \in \mathcal{D}\label{eq_sto_lin_2}\\
&V_{i,d} + \Delta X_{i,d} \leq \Delta + A_{i,d}, &\forall d \in \mathcal{D},\forall i \in \mathcal{C} \label{eq_sto_lin_3}
\end{align}

\begin{align}
V_{i,d} = \left( \prod_{j \in \mathcal{C}, j < i}\hspace{-0.5em}( 1 -\mathbb{P}) \right) \mathbb{P} C^{(p)} \left( N_d - (\hspace{-0.5em}\sum_{j \in \mathcal{C}, j <i}\hspace{-0.8em}1) +1\right), \nonumber\\ \forall d \in \mathcal{D}, i \in \mathcal{C} \label{eq_sto_lin_4}
\end{align}

The constraints in (\ref{eq_sto_lin_1}) and (\ref{eq_sto_lin_2}) ensure that the customer serving order does not exceed the total number of the customers. For example, if drone $d$ serves five customers, i.e., $N_d = 5$, $X_{i,d} = 1$ for $i = 1,\ldots,5$, and $X_{i,d}=0$ for $i > 5$. Note that $X_{i,d}$ is not related to a specific customer, and the set $\mathcal{C}$ and the index $i$ are used for simplifying the parameter declaration. The constraint in (\ref{eq_sto_lin_3}) ensures that $A_{i,d} = V_{i,d}$ when $i$ is the first $N_d$ customers, i.e., $X_{i,d} =1$ for $i=1,\ldots,N_d$. The constraint in (\ref{eq_sto_lin_4}) calculates the expected value of the penalty incurred by the drone breakdown. 

The above optimization can be solved by using a linear programming solver. $W_d^*$, $Y_{i,d,p}^*$, $Z_i^*$, $T_p^*$, $M_{i,p,q}^*$, $B_{d,p}^*$, $N_d^*$, $A_{i,d}^*$, $X_{i,d}^*$, and $V_{i,d}^*$ represent the optimal solutions of $W_d$, $Y_{i,d,p}$, $Z_i$, $T_p$, $M_{i,p,q}$, $B_{d,p}$, $N_d$, $A_{i,d}$, $X_{i,d}$, and $V_{i,d}$, respectively. The optimal total delivery cost is denoted by $\mathbb{O}^{\mathrm{ST}*}$ which is used in the cost management.

\section{Cost Management}
\label{sec_sys_cost}

With the solution of the package assignment, the delivery cost will be shared among the multiple cooperative shippers in the same coalition. Let $\mathbb{O}^{\mathrm{ST}*} (\mathcal{S})$ denote the total delivery cost incurred to the cooperative suppliers in coalition $\mathcal{S}$. Note that $\mathbb{O}^{\mathrm{ST}*} (\mathcal{S})$ can be obtained from solving the package assignment problem, e.g., the solution of (\ref{eq_sto_obj}), the detail of which is given in Section~\ref{sec_package}. Let $v_p(\mathcal{S})$ denote the cost that shipper $p$ needs to pay for the delivery when joining coalition $\mathcal{S}$. Thus, $\mathbb{O} ^{\mathrm{ST}*}(\mathcal{S}) = \sum_{p \in \mathcal{S}}v_p( \mathcal{S} )$. In this paper, we use the Shapley value~\cite{ref_shapley} to achieve the solution of the cost management, i.e., the delivery cost shared by the cooperative shippers in the same coalition. The Shapley value can be expressed as follows:
\begin{align}
v_p(\mathcal{S})= \hspace{-1.5em}\sum_{\mathcal{Q} \subseteq \mathcal{S} \setminus \{p\}}\hspace{-1em}\frac{|\mathcal{Q}|!(|\mathcal{S}|- |\mathcal{Q}| -1)!}{|\mathcal{S}|!}\left( \mathbb{O}^{\mathrm{ST}*} ( \mathcal{Q}\cup \{p\} ) - \mathbb{O} ^{\mathrm{ST}*}(\mathcal{Q}) \right), 
\label{eq_shapley}
\end{align}
where $|\mathcal{P}|$ represents the total number of shippers, and $|\mathcal{S}|$ represents the number of shippers in coalition $\mathcal{S}$. 
The four properties of Shapley value are listed as follow: 
\begin{itemize}

\item\textbf{Efficiency} The coalition cost is equal to the summation of the individual costs, i.e., $\mathbb{O} ^{\mathrm{ST}*}(\mathcal{S}) = \sum_{p \in \mathcal{S}}v_p( \mathcal{S} )$. As a result, the sum of the coalition cost will be minimized.
\item\textbf{Symmetry} If shippers $p$ and $q$ join a coalition $\mathcal{S}$, and the total coalition cost is the same, i.e., $\mathbb{O}^{\mathrm{ST}*}( \mathcal{S} \cup \{p\}) = \mathbb{O}^{\mathrm{ST}*}(\mathcal{S} \cup \{q\})$, then 
$v_p(\mathcal{S})=v_q(\mathcal{S})$. As such, the shared cost for shipper $p$ and shipper $q$ will be the same if they join coalition $\mathcal{S}$ and contribute the same. 
\item\textbf{Linearity} Let $\mathbb{O}^{\mathrm{A}}$ and $\mathbb{O}^{\mathrm{B}}$ be a function, which is similar to function $\mathbb{O}^{\mathrm{ST}}$. When $\mathbb{O}^{\mathrm{A}}$ and $\mathbb{O}^{\mathrm{B}}$ are characteristic functions, 
 $v_p(\mathcal{S}, \mathbb{O}^{\mathrm{A}} + \mathbb{O}^{\mathrm{B}})= v_p(\mathcal{S}, \mathbb{O}^{\mathrm{B}} + \mathbb{O}^{\mathrm{A}}) = v_p(\mathcal{S}, \mathbb{O}^{\mathrm{A}}) + v_p(\mathcal{S}, \mathbb{O}^{\mathrm{B}})$. Note that $v_p(\mathcal{S}, \mathbb{O}^{\mathrm{ST}})$ is referred to $v_p(\mathcal{S})$  since we only use  function $\mathbb{O}^{\mathrm{ST}}$ in this paper.
\item\textbf{Zero player} A shipper $p$ does not need to pay the delivery cost if the coalition cost remains the same when shipper $p$ joins in, i.e., $v_p(\mathcal{S}) = 0$ when $ \mathbb{O}^{\mathrm{ST}*}(\mathcal{Q}) =  \mathbb{O}^{\mathrm{ST}*}(\mathcal{Q}\cup\{p\})$ and $\mathcal{Q} \subseteq \mathcal{S}$.
\end{itemize}

\section{Cooperation Management: Static Bayesian Coalition Formation Game}
\label{sec_static_game}

The shippers can partner, i.e., cooperate, with each other and form a coalition for establishing a resource pool for package delivery. In this regard, all shippers need to decide to cooperate or not cooperate with other shippers to minimize their costs. Recall that $\mathcal{P}$ is a set of all shippers in the system, and $\mathcal{S}$ is a coalition, i.e., a set, of cooperative shippers establishing a resource pool. Let $\Phi = \{\phi_1,\phi_2,\dots,\phi_{|\Phi|} \}$ denote the set of all possible coalition structures, where $|\Phi|$ is the total number of all coalition structures which can be calculated from Bell number~\cite{ref_shapley} with $|\mathcal{P}|$ as a parameter. In a coalition structure, there is one or more coalition, i.e., $\phi = \{\dots,\mathcal{S}_l,\dots\}$. We have $\mathcal{S}_l \cap \mathcal{S}_{l'} = \emptyset$ for all $l\neq l'$ when coalitions $\mathcal{S}_l$ and $\mathcal{S}_{l'}$ are in the same coalition structure $\phi$. $\bigcup_{l \in \phi}\mathcal{S}_l = \mathcal{P}$ for every coalition structure. The coalition structure is basically a set of all coalitions that include all the shippers, i.e., ${\mathcal{P}} = \mathcal{S}_1 \cup \mathcal{S}_2 \cup \cdots \cup \mathcal{S}_{|\phi|}$. For example, with $\mathcal{P} = \{p_1,p_2,p_3, p_4\}$ and $\phi = \{\{p_1,p_2\}, \{p_3,p_4\}\}$, this coalition structure consists of two coalitions, i.e., $\mathcal{S}_1 = \{p_1,p_2\}$ and $\mathcal{S}_2 = \{p_3,p_4\}$, where the shipper $p_1$ cooperates with the shipper $p_2$ and the shipper $p_3$ cooperates with the shipper $p_4$. 

Next, we consider the fact that shippers can misbehave when they join a coalition. For example, given the solution of the package assignment for a coalition, a misbehaving shipper may not deliver packages of  other shippers in the same coalition, even if they agree to cooperate. Therefore, we classify shippers into two types as follows: 
\begin{itemize}
\item \textbf{Good} (${\mathrm{T^G}}$) is a shipper that delivers all assigned packages. 
\item \textbf{Bad} (${\mathrm{T^B}}$) is a shipper that does not deliver all assigned packages, and the undelivered packages belong to other shippers in the same coalition.
\end{itemize}

Each shipper has a belief probability about other shippers. Let $b_{pq}$ denote a type that shipper $p$ believes about shipper $q$. Let $P(b_{pq} = {\mathrm{T^G}})$ and $P(b_{pq} = {\mathrm{T^B}})$ denote the probabilities that shipper $p$ believes that shipper $q$ is a good cooperative shipper and a bad cooperative shipper, respectively. Moreover, shipper $p$ believes that shipper $q$ is a good cooperative shipper with probability $P(b_{pq} = {\mathrm{T^G}})= \lambda_{pq}$. Thus, $P(b_{pq} = {\mathrm{T^B}})= 1 - \lambda_{pq}$. 


To address the uncertainty in the suppleirs' behavior, we formulate the BCoSDD as a Bayesian coalition formation game as presented as follows:
\begin{align}
G = <\mathcal{S},\mathbb{T},P, (\mu_p)_{p \in S}, (\succeq_p)_{p \in S}>, 
\label{eq_game}
\end{align}
where the definitions of the parameters are given as follows.
\begin{itemize}
\item Let $\mathcal{S} = \{\dots, P_p, \dots\}$ denote a coalition, i.e., a set of shippers, that join the resource pool, and $\mathcal{S} \subset \mathcal{P}$. $|\mathcal{S}|$ denotes the total number of the shippers in coalition $\mathcal{S}$. Note that $p$ and $q$ are the indexes of the elements in set $\mathcal{S}$.

\item Let $\mathbb{T} = \prod_{p \in \mathcal{S}} \mathbb{T}_p $ be the type space of all shippers in $\mathcal{S}$, where $\mathbb{T}_p = \{ {\mathrm{T^G}}, {\mathrm{T^B}}\}$ denotes a set of possible types of shipper $p$.

\item $P$ is the probability distribution over the types of shippers, i.e., $P(b_{pq} = {\mathrm{T^B}})$ and $P(b_{pq} = {\mathrm{T^G}})$. 

\item Let $\mu_p$ denote the expected payoff, i.e., the delivery cost that shipper $p$ needs to pay. 
\item Let $\succeq_p$ present the preference of shipper $p$. 
\end{itemize}

In the following, we give the details of the expected payoff and the preference condition. Then, we present the methods to achieve the solution of the Bayesian coalition formation game, which can be referred to as the stable coalition structure. We can analyze the stable coalition structure by using the markov transition process. We can also reach the stable coalition structure by using the merge and split algorithm.

\subsection{Expected Payoff ($\mu_p$)}
\label{sec_game_mu}

If we do not consider the uncertainty of misbehaving shippers, we can directly use the Shapley value, which is discussed in Section~\ref{sec_sys_cost}. Since we consider the misbehaving uncertainty, the expected payoff has to be derived. Again, $\mu_p$ denotes the expected delivery cost that shipper $p$ needs to pay, including the penalty of the undelivered packages which is incurred from the misbehaving shipper. Let $K_p(\mathcal{S}) = \prod_{q \in \mathcal{S}\setminus \{p\}}\mathbb{T}_q$ denote the set of the possible belief combinations between shipper $p$ and the other cooperative shippers, i.e. $q \in \mathcal{S}\setminus \{p\}$. The total number of members in $K_p(S)$ is $2^{(|S|-1)}$. For example, if $\mathcal{S}=\{p_1,p_2,p_3\}$ and $p=p_1$, $K_1(\mathcal{S}) = \{k_1,k_2,\dots,k_{2^{|s|-1}}\} = \mathbb{T}_2 \times \mathbb{T}_3 = \{k_1 = (b_{12} = {\mathrm{T^G}}, b_{13} ={\mathrm{T^G}}), k_2 = (b_{12} = {\mathrm{T^G}}, b_{13} ={\mathrm{T^B}}), k_3 = (b_{12} ={\mathrm{T^B}},b_{13}={\mathrm{T^G}}), k_4 = (b_{12}={\mathrm{T^B}},b_{13}={\mathrm{T^B}})\}$. The expected delivery cost of shipper $p$ can be calculated from 
\begin{align}
&\mu_p(\mathcal{S}) = v_p(\mathcal{S})+ \sum_{k \in K_p(\mathcal{S})} \sum_{q\in S\setminus \{p\}}\Omega_{kp}\Lambda_{kpq}, 
\label{bay_mu}
\end{align}
\noindent where $\Omega_{kp}$ is the probability that the belief combination $k$ happens, and $\Lambda_{kpq}$ denotes the incurred cost when the belief combination $k$ happens, and shipper $q$ does not delivery packages for shipper $p$. Again, $v_p(\mathcal{S})$ can be obtained from the cost management in Section~\ref{sec_sys_cost}. $\Omega_{kp}$ is obtained from
\begin{align}
&\Omega_{kp} = \prod_{q \in \mathcal{S}\setminus \{p\}} P(b_{pq} = t), \label{bay_exp_1}
\end{align}
where $t$ is the type taken from space $\mathbb{T}$, and $\Lambda_{kpq}$ is obtained from
\begin{align}
&\Lambda_{kpq} = 
\begin{cases} 0, & \mbox{if } b_{pq} = {\mathrm{T^G}}, \\ \theta_{pq} C^{(p)}(1-\lambda_{pq}), & \mbox{if } b_{pq} = {\mathrm{T^B}}. \end{cases}
\label{bay_exp_2}
\end{align}
\noindent $\theta_{pq} $ denotes the total number of the packages that shipper $p$ transfers to shipper $q$, i.e., $\theta_{pq}= \sum_{i \in \mathcal{C}}M_{p,q}^*$, where $M_{p,q}^*$ can be obtained from the package assignment in Section~\ref{sec_package}. 

\subsection{Preference Relations}
\label{sec_bay_perfer}
We adopt the preference relations from~\cite{ref_merge}. Again, let $\succeq_p$ present the preference of shipper $p$. For example, $(\mathcal{S}_1,\phi_m) \succeq_p (\mathcal{S}_2,\phi_n)$ means that shipper $p$ prefers to join coalition $\mathcal{S}_1$ from coalition structure $\phi_m$ over to join coalition $\mathcal{S}_2$ from coalition structure $\phi_n$. The preference relation of a shipper can be calculated from
\begin{align}
(\mathcal{S}_1,\phi_m) \succeq_p (\mathcal{S}_2,\phi_n) \longleftrightarrow \nu_p(\mathcal{S}_1,\phi_m) \geq \nu_p(\mathcal{S}_2,\phi_n)	, \label{bay_prefer_1}	
\end{align}

\noindent where $p \in \mathcal{S}_1$, $p \in \mathcal{S}_2$, and $\mathcal{S}_1$ and $\mathcal{S}_2$ are any two coalitions. The preference function is represented as $\nu_p$ and is defined as follows:
\begin{align}
\nu_p(\mathcal{S},\phi) = 	\left\lbrace \begin{array}{ll}
\mu_p(\mathcal{S},\phi), &\textbf{if } \mu_q(\mathcal{S},\phi) \leq \mu_q(\mathcal{S}\setminus \{q\},\phi ) \\
& \forall q \in \mathcal{S}\setminus \{q\} \text{ and }|\mathcal{S}|\neq 1 \\
\\ \\
0& \textbf{otherwise},
\end{array} \right.
\label{bay_prefer_2}
\end{align}
This function requires three inputs, i.e., (i) a main shipper $p$, (ii) coalition $\mathcal{S}$, and (iii) the coalition structure of coalition $\mathcal{S}$, which is $\phi$. From (\ref{bay_prefer_2}), shipper $p$ will switch from the current coalition $\mathcal{S}_m$ to a another coalition $\mathcal{S}_n $ when shipper $p$ prefers the new coalition $\mathcal{S}_n$ over coalition $\mathcal{S}_m$, i.e., $(\mathcal{S}_n \cup \{p\},\phi_n) \succeq_p (\mathcal{S}_m,\phi)$, where $\mathcal{S}_m$ and $\mathcal{S}_n \in \phi$. As a result, the new coalition structure is $\phi_n= \{\phi\setminus\{\mathcal{S}_m,\mathcal{S}_n\} \} \cup \{\mathcal{S}_m \setminus \{p\},\mathcal{S}_n \cup \{p\}\}$. 

\subsection{Merge and Split Algorithm}
According to the coalition structure switching, one shipper can join or leave a coalition at a time. Therefore, the merge and split algorithm is adopted from~\cite{ref_merge} as presented in Algorithm~\ref{bay_algo}. The algorithm is composed of four major steps, i.e., neighboring coalition search (Line 3), the calculation of Shapley value (Line 8 to Line 9), the calculation of Bayesian cost (Line 12 to Line 17), and stable state checking (Line 18 to Line 21). The algorithm terminates when the current coalition structure $\phi_m$ cannot change to any neighboring coalition structure. 

\begin{algorithm}[]
 \caption{Merge and Split Algorithm for Cooperation Management}
\label{bay_algo}
 \begin{algorithmic}[1]
 
 \renewcommand{\algorithmicrequire}{\textbf{Input:}}
 \renewcommand{\algorithmicensure}{\textbf{Output:}}
 \REQUIRE Set $\mathcal{P}$ and all the coalition cost $V(\mathcal{S})$ obtained by solving the package assignment optimization problem
 \ENSURE A stable coalition structure, the individual cost for each shipper, the drone delivery planning\\
 \textit{Initialization} : All shippers do not cooperate

 \WHILE { $\phi_m$ changes}
 \STATE $\phi_n$ = The list of $\phi_m$ neighboring coalition structures, which can be found by using the neighboring coalition discovery algorithm~\cite{ref_merge}, \cite{ref_neighbor}. 
 \FOR {every $b \in \phi_n$}
 \FOR {every $p \in \mathcal{P}$}
 
 \STATE $\mathcal{C}_o$ and $\mathcal{C}_b$ are the coalitions that $p$ belongs to coalition structure $\phi_m$ and $\phi_n$, respectively.
 \FOR {every $S \in \phi_m$}
 \STATE Obtain $\mathbb{O}^{ST*}(\mathcal{S})$, $\forall \mathcal{S}\in \phi_m$ { \em //call the coalition cost, which can be obtained by solving the package assignment optimization problem of coalition $S$}\\
 $v_p(S) \leftarrow \mathbb{O}^{ST*}(\mathcal{S})$ {\em //solve for the Shapley Value}
 \ENDFOR
 \FOR {every $S \in \phi_n$}
 \STATE Obtain $\mathbb{O}^{ST*}(\mathcal{S})$, $\forall \mathcal{S}\in \phi_n$ { \em //call the coalition cost, which can be obtained by solving the package assignment optimization problem of coalition $S$}\\
 $v_p(S) \leftarrow \mathbb{O}^{ST*}(\mathcal{S})$ {\em //solve for the Shapley Value}
 \ENDFOR
\IF { $p$ has visited $\mathcal{C}_b$ before \textbf{and} $|\mathcal{C}_b| > 1 $ }
\STATE $\nu_p(\mathcal{C}_b,\phi_n) =0$
\ELSE
\STATE $\nu_p(\mathcal{C}_b,\phi_n)$ is calculated from~(\ref{bay_prefer_2})
\ENDIF
 \STATE Calculate $\nu_p(\mathcal{C}_o,\phi_m)$ from~(\ref{bay_prefer_2})
 \IF {$\nu_p(\mathcal{C}_o,\phi_m) \geq \nu_p(\mathcal{C}_b,\phi_n)$}
 \STATE $\phi_m = \phi_n$; {\em // the coalition structure has changed}
 \textbf{break}; {\em// break to the while loop (Line 2)}
 \ENDIF \ENDFOR
 \ENDFOR
 \ENDWHILE
 \end{algorithmic}

 \end{algorithm}

\subsection{Markov Transition Process}

\begin{figure}
\includegraphics[width=0.5\textwidth]{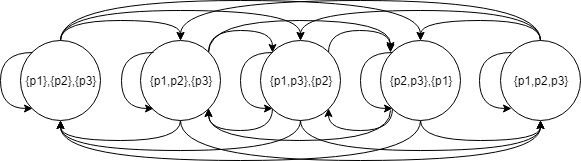}
\caption{The state transition diagram of three shippers.}
\label{fig_dyn_mtp}
\end{figure}

Let a set of coalition structures, i.e., $\Phi = \{\phi_1,\phi_2,\dots,\phi_{|\Phi|}\}$, be the state space of the markov transition process. Let $\rho_{\phi_m,\phi_n}$ denote a transition probability when transiting from state $\phi_m$ to state $\phi_n$, which can be referred to as coalition structure $\phi_m$ is changed to coalition structure $\phi_n$. Based on the preference relations in Sectione~\ref{sec_bay_perfer}, only one shipper can change from coalition structure $\phi_m$ to coalition struture $\phi_n$ at a time. Figure~\ref{fig_dyn_mtp} shows the example of the state transition diagram when there are three shippers. Let $\mathcal{H}_\phi$ denote the set of the neighboring coalition structures of coalition structure $\phi$, which only one shipper can change a coalition at a time. For example, if $\phi_m = \{\{p_1\},\{p_2\},\{p_3\}\}$, $\mathcal{H}_{\phi_m} = \{\{\{p_1,p_2\},\{p_3\}\}, \{\{p_1,p_3\},\{p_2\}\}, \{\{p_2,p_3\},\{p_1\}\}\}$. Let $\mathbb{B}_{\phi_m,\phi_n}$ denote the probability that coalition structure $\phi_m$ will change to coalition structure $\phi_n$, which can be calculated from
\begin{align}
&\mathbb{B}_{\phi_m,\phi_n} = \left\lbrace \begin{array}{ll}
1 - \sum_{ \phi \in \Phi} \mathbb{B}_{\phi_m,\phi}, & \textbf{if } \phi_m=\phi_n,\\\\
0,& \textbf{else if } \phi_n \notin \mathcal{H}_{\phi_m}, \\\\
\varepsilon \beta_{\phi_m,\phi_n},& \textbf{else if } \\
& \nu_b(S_m,\phi_m) < \nu_b(S_n,\phi_n), \\
&\text{for all } b \in \mathcal{B}_{\phi_m,\phi_n}\\\\
 (1-\varepsilon)\beta_{\phi_m,\phi_n},& \textbf{otherwise},
\end{array}\right. \label{bay_mtp_1}\\
&\beta_{\phi_m,\phi_n} = \alpha^{|\mathcal{B}_{\phi_m,\phi_n}|}(1-\alpha)^{|\mathcal{P}|-|\mathcal{B}_{\phi_m,\phi_n}|}
\end{align}
where $\mathcal{B}_{\phi_m,\phi_n}$ represents the set of customers that are involved to the transition from coalition structure $\phi_m$ to coalition structure $\phi_n$, and $|\mathcal{B}_{\phi_m,\phi_n}|$ denotes the total number of the shippers in set $\mathcal{B}_{\phi_m,\phi_n}$. Note that $\alpha$ denotes the probability that a shipper changes its coalition. 

Given transition probability matrix {\boldmath Q}, which is expressed as follows:
\begin{align}
\text{\boldmath{Q}} = 
\begin{bmatrix} 
\mathbb{B}_{1,1} & \mathbb{B}_{1,2} &\dots & \mathbb{B}_{1,|\Phi|} \\
\mathbb{B}_{2,1} & \mathbb{B}_{2,2} & \dots& \mathbb{B}_{2,|\Phi|} \\
\vdots & \vdots & \ddots & \\
\mathbb{B}_{|\Phi|,1} & \mathbb{B}_{|\Phi|,2} & & \mathbb{B}_{|\Phi|,|\Phi|}
\end{bmatrix}	,
\label{bay_q_matrix}
\end{align}
we can obtain the stationary probability vector by solving
\begin{align}
	\vec{\Pi}^\top \text{\boldmath{Q}} =\vec{\Pi}^\top,	\quad	\vec{\Pi}^\top \vec{1} = 1	,
\label{bay_mtp_2}
\end{align}
\noindent where $\vec{\Pi} = [\pi_1,\pi_2,\dots,\pi_{|\Phi|}]^\top$ denotes the stationary probability vector, and $\pi_{\phi}$ is the probability that coalition structure $\phi$ will be formed. 


\section{Cooperation Management: Dynamic Bayesian Coalition Formation Game}
\label{sec_dynamic_game}
As discussed in the previous section, we can use the merge and split algorithm to obtain a stable coalition structure in the static Bayesian coalition formation game. In this section, we extend the static Bayesian coalition formation game to the dynamic one that allows a shipper to observe behaviors and update the belief about the other cooperative shippers in the same coalition continuously. Therefore, the dynamic Bayesian coalition formation game can be used even when the shippers do not have prior knowledge about the other shippers. In particular, a shipper observes the behavior of the cooperative shippers whether all the packages are successfully delivered by them or not. The belief probability, i.e., $P(b_{p,q} = {\mathrm{T^G}})$ and $P(b_{p,q} = {\mathrm{T^B}})$, is updated based on the observation.

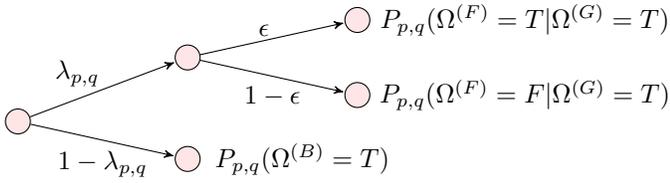
\begin{figure}
\begin{tikzpicture}[->,>=stealth',shorten >=0.5pt,auto,node distance=1.7cm,
 main node/.style={circle,fill=red!10,draw, minimum size=0.5mm}]
 \node[main node] (D5) {};
 \node[main node] (D1) [below left of=D5, yshift=2em, xshift = -3em] {};
 \node[main node] (D2) [below left of=D1, yshift=1em, xshift = -3em] {};
 \node[main node] (D3) [below right of=D1, yshift=2em, xshift = 3em] {};
 \node[main node] (D4) [below left of=D3, yshift=1em, xshift = -3em] {};
 \node [right of=D5, xshift = 1.5em] {$P_{p,q}(\Omega^{(F)}=T|\Omega^{(G)}=T)$};
 \node [right of=D3, xshift = 1.5em] {$P_{p,q}(\Omega^{(F)}=F|\Omega^{(G)}=T)$};
 \node [right of=D4, xshift = -0.5em] {$P_{p,q}(\Omega^{(B)}=T)$};
 \path[every node/.style={inner sep=1pt}]
 (D1) edge node {$\epsilon$} (D5) 
(D2) edge node {$\lambda_{p,q}$} (D1)
(D2) edge node [below=1mm] {$1-\lambda_{p,q}$} (D4)
(D1) edge node [below=1mm] {$1-\epsilon$} (D3);
\end{tikzpicture} 
\caption{Tree diagram of the belief update mechanism of shipper $p$ about shipper $q$.}
\label{fig_dyn_tree}
\end{figure}

We propose the belief update mechanism based on the Bayes' theorem~\cite{ref_baye} and the algorithm in~\cite{ref_aj.mink}. There are three possible conditions involved in the belief update mechanism of shipper $p$ about shipper $q$, namely: (i) $\omega^{(F)}$ is the condition that shipper q's delivery is not successful due to technical issues, (ii) $\omega^{(G)}$ is the condition that shipper q is a cooperative shipper., and (iii) $\Omega^{(B)}$ is the condition that shipper $q$ is a bad cooperative shipper, i.e., its real type is {\bf Bad}. Let $\Omega=T$ represent that the condition $\Omega$ is true, and $\Omega=F$ otherwise, where $\Omega \in \{\Omega^{(F)}, \Omega^{(G)} , \Omega^{(B)} \}$. Here, we have $P_{p,q}(\Omega^{(B)} =T) = P_{p,q}(\Omega^{(G)} =F) = 1 - P_{p,q}(\Omega^{(G)}=T)$. The probability that the condition $\Omega^{(G)}$ is true is equal to the belief probability, i.e., $P_{p,q}(\Omega^{(G)}) = P(b_{p,q}={\mathrm{T^G}})=\lambda_{p,q}$. Figure~\ref{fig_dyn_tree} presents the tree diagram of the belief update mechanism with the three conditions and their corresponding probabilities. Let $\epsilon$ denote an error probability that the delivery by the good cooperative shipper is unsuccessful.

Likewise, Table~\ref{bay_table} shows all the joint probabilities. There are eight cases, but only three cases can happen. The cases $\omega_1$, $ \omega_4$, $\omega_5$ and $\omega_8$ will not happen because $\Omega^{(G)}$ and $\Omega^{(B)}$ cannot be true at the same time, and $\omega_3$ will not happen because the bad cooperative shipper will not deliver the package. From Table~\ref{bay_table}, we can show that
\begin{align}
&P_{p,q}(\Omega^{(F)}=T|\Omega^{(G)}=T) + \nonumber \\
&P_{p,q}(\Omega^{(F)}=F|\Omega^{(G)}=T) + P_{p,q}(\Omega^{(B)}=T) = 1,
\label{eq_bay_first}
\end{align}
where
\begin{align}
 & P_{p,q}(\Omega^{(F)}=T|\Omega^{(G)}=T) = \lambda_{p,q}\epsilon, \\
 & P_{p,q}(\Omega^{(F)}=F|\Omega^{(G)}=T) = \lambda_{p,q}(1-\epsilon), \\
 & P_{p,q}(\Omega^{(B)}=T) = 1-\lambda_{p,q}.
\end{align}
Consequently, the probabilities that a package of shipper $p$ is delivered successfully and unsuccessfully by shipper $q$, i.e., $P_{p,q}(S)$ and $P_{p,q}(U)$, can be calculated from 
\begin{align}
&P_{p,q}(S) &=& P_{p,q}(\Omega^{(F)}=F|\Omega^{(G)}=T) = \dfrac{\theta_{p,q}'}{\theta_{p,q}}\label{eq_bay_suc}\\
&P_{p,q}(U) &=& P_{p,q}(\Omega^{(F)}=T|\Omega^{(G)}=T) + P_{p,q}(\Omega^{(B)}=T) \nonumber\\
&&=& 1- \dfrac{\theta_{p,q}'}{\theta_{p,q}} \label{eq_bay_unsec},
\end{align}

\noindent where $\theta_{p,q}'$ denotes the exact number of the packages that shipper $q$ successfully delivers for shipper $p$. Again, $\theta_{p,q}$ represents the total number of the packages that shipper $p$ transfers to shipper $q$, and shipper $q$ needs to deliver them. From (\ref{eq_bay_suc}) and (\ref{eq_bay_unsec}), we can obtain the new belief probability $\lambda_{p,q}^{t+1}$ in iteration $t$ based on the observation as follows:
\begin{align}
\lambda_{p,q}^{t+1} = \left\lbrace \begin{array}{ll}
\dfrac{\theta_{p,q}'}{\theta_{p,q}(1-\epsilon)}, & \textbf{if }\dfrac{\theta_{p,q}'}{\theta_{p,q}} < 1 - \epsilon\\\\
1, & \textbf{otherwise}.
\end{array}\right.
\label{eq_bay_lambda}
\end{align}
 However, the new belief probability $\lambda_{p,q}$ from (\ref{eq_bay_lambda}) is independent of the past as the probability is based on only current observation. Therefore, the weight adjustment from the exponential moving average (EMA) method~\cite{ref_ema} can be used to update the belief probability as follows:
\begin{align}
\lambda_{p,q}^{t+1} = w_1\lambda_{p,q}^{t} + w_2\lambda_{p,q}^{t+1}
\label{eq_bay_weight}
\end{align} 
where $\lambda_{p,q}^{t+1}$ on the right hand side is obtained from (\ref{eq_bay_lambda}). The constants $w_1$ and $w_2$ are adjustable weight parameters, for $w_1+w_2 =1$ and $0 \leq w_1,w_2 \leq 1$.
 
\begin{table}[t]
\caption{Truth table to the joint probabilities of the belief update mechanism. }
\label{bay_table}
\normalsize \centering
\begin{tabular}{|c|ccc|c|}
\hline
 Scenario  & $\Omega^{(F)}$ & $\Omega^{(G)}$ & $\Omega^{(B)}$ & Joint Probability   \\\hline 
$\omega_1$ & $T$ & $T$ & $T$ & 0 \\
$\omega_2$ & $T$ & $T$ & $F$ & $\lambda_{p,q}\epsilon$  \\
$\omega_3$ & $T$ & $F$ & $T$ & 0 \\
$\omega_4$ & $T$ & $F$ & $F$ & 0 \\
$\omega_5$ & $F$ & $T$ & $T$ & 0 \\
$\omega_6$ & $F$ & $T$ & $F$ & $\lambda_{p,q}(1-\epsilon)$ \\
$\omega_7$ & $F$ & $F$ & $T$ & $1- \lambda_{p,q}$   \\
$\omega_8$ & $F$ & $F$ & $F$ & 0  \\\hline      
\end{tabular}
\end{table}

\begin{algorithm}[t]
 \caption{ Dynamic Bayesian Coalition Formation }
\label{bay_algo_dynamic}
 \begin{algorithmic}[1]
 
 \renewcommand{\algorithmicrequire}{\textbf{Input:}}
 \renewcommand{\algorithmicensure}{\textbf{Output:}}
 \ENSURE Drone delivery planning solution\\

 \REPEAT
 \STATE Perform Algorithm~\ref{bay_algo}
 \STATE Perform package delivery based on the solution of Algorithm~\ref{bay_algo}
 \STATE Observe whether the packages are successfully delivered or not, i.e., $\theta_{p,q}'$. 
 \FOR {every $p \in \mathcal{P}$}
 \FOR {every $q \in \mathcal{P}$}
	\IF{shipper $p$ transfers packages to shipper $q$}
	\STATE Update the belief probability as in (\ref{eq_bay_weight})
	\ELSE
	\STATE $\lambda^{t+1}_{p,q} \leftarrow \lambda^{t}_{p,q}$
	\ENDIF
 \ENDFOR
 \ENDFOR
 \STATE $t \leftarrow t+1$ 
 \UNTIL{Terminates}
 \end{algorithmic}
 \end{algorithm}

Algorithm~\ref{bay_algo_dynamic} shows all steps of dynamic Bayesian coalition formation. The belief update mechanism, which is based on (\ref{eq_bay_first}) to (\ref{eq_bay_weight}) and the merge and split algorithm are used in Algorithm~\ref{bay_algo_dynamic}. When the shippers perform the package delivery by following the planning solution for multiple times, the belief probabilities converge to constants, and thus the algorithm also converges to the same stable coalition structure as achieved when all the types of shippers are exactly known. 


\section{Performance Evaluation}
\label{sec_experiment}
In this section, we first describe the experimental setup. Next, we discuss the experimental results to demonstrate the performance of the BCoSDD framework.

\begin{figure*}
\begin{subfigure}[t]{0.5\textwidth}
\includegraphics[width=\textwidth, height = 22em]{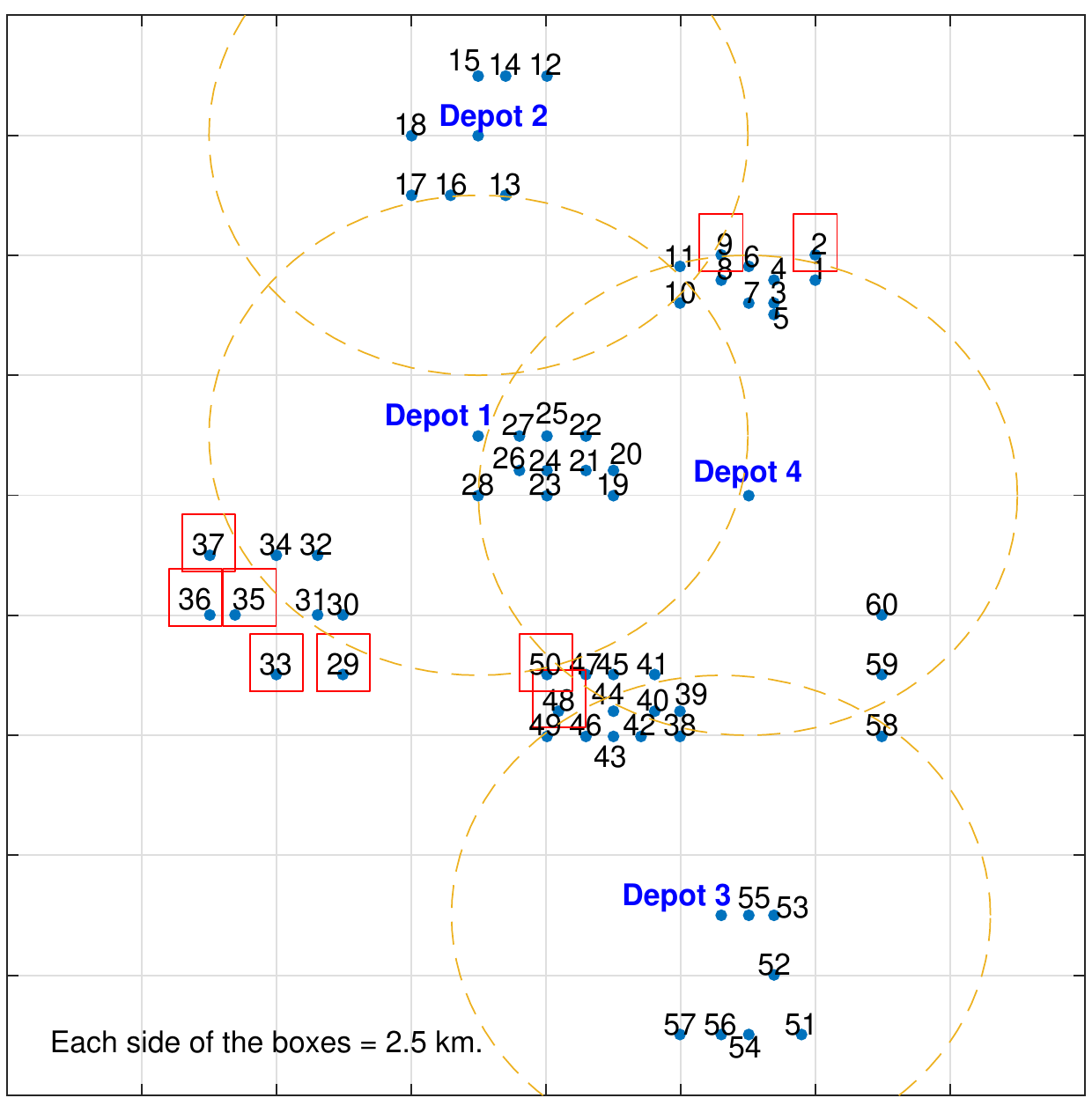}
\caption{The synthesized Solomon Benchmark data.} \label{fig_map_solomon}
\end{subfigure} 
\begin{subfigure}[t]{0.5\textwidth}
\includegraphics[width=\textwidth, height = 21.5em]{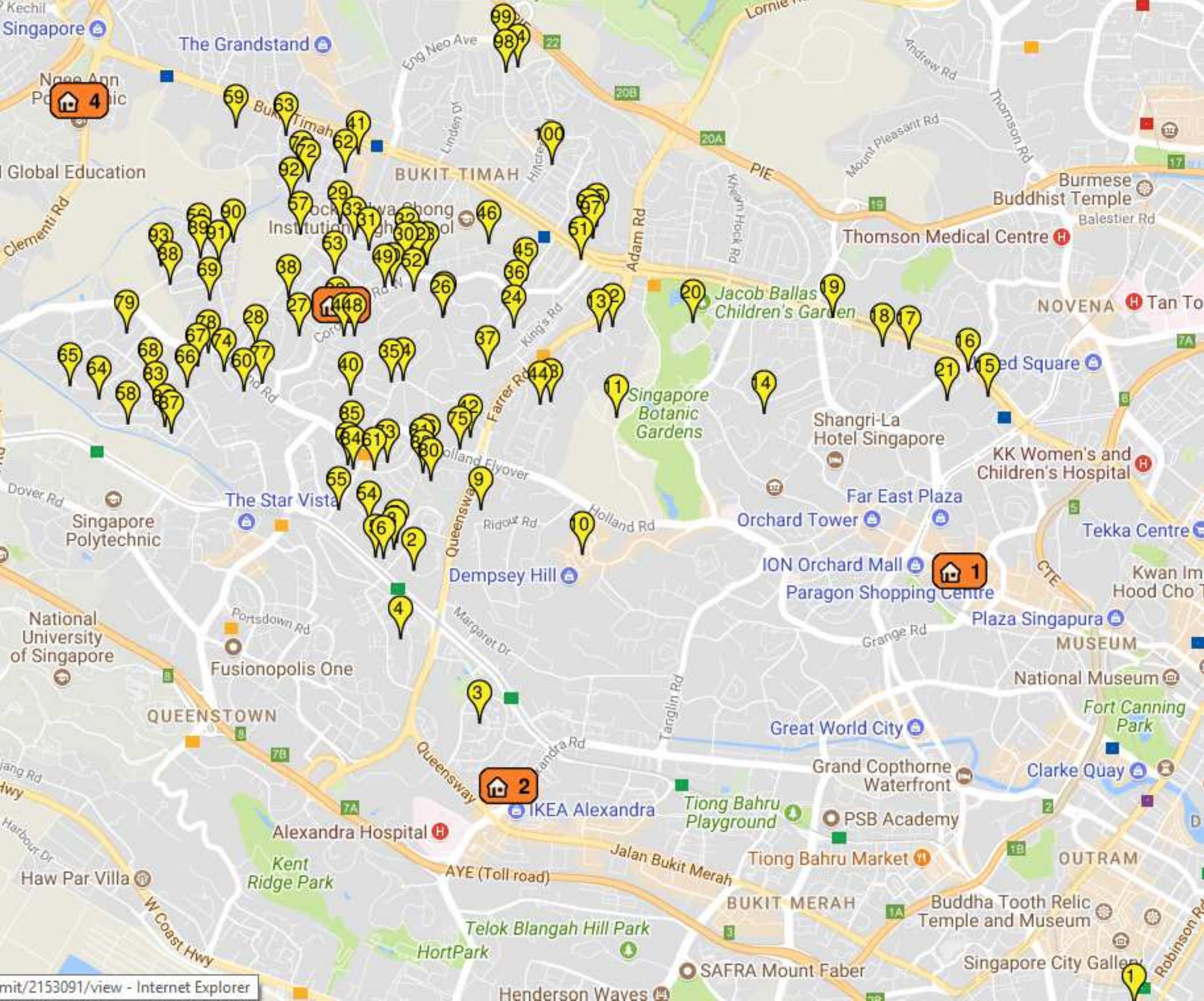} 
\caption{A real data from logistics industry in Singapore.} \label{fig_map_real}
\end{subfigure} 
\caption{The locations of customers and depots.}
\end{figure*}

\subsection{Experiment Setup}

We consider four shippers, i.e., $\mathcal{P} = \{p_1,p_2,p_3,p_4\}$. Each shipper $p \in \mathcal{P}$ has one drone, and their drones are identical. The initial cost of the drones is set as $C^{(i)} = S\$100$. The limits of the drones are set as $f_d = 5$ kilograms, $e_d = 10$ kilometers, $l_d = 150$ kilometers, and $h_d = 8$ hours. We use the average flying speed of drones which is $s_d = 30$ kilometers per hour in the experiments. The serving time is set as $ t_i= 15$ minutes for all $i \in \mathcal{C}$. All the customer packages in the experiments weight less than 5 kilograms, i.e., $a_i \leq 5$ kilograms. Note that the parameter setting is based on~\cite{maggie_TITS} and a Singapore company. When the shippers cooperate, they share their customers and drones in the resource pool. The transferring cost from the original depot to the new depot is set as $C^{(t)} = S\$30$ per transfer. The outsourcing cost is set as $C^{(c)}_i = S\$16$ based on the speedpost service provided by Singpost company~\cite{ref_singpost}. Similarly, the penalty of breakdown is set as $C^{(p)}_i = S\$16$. Therefore, the shippers may handle the unsuccessful delivery packages by outsourcing them to the carrier. The probability that shipper $p$ believes that shipper $q$ are a good cooperative shipper is set as $P(b_{pq} = {\mathrm{T^G}}) = 0.9$ for all $p,q \in \mathcal{P}$.

We evaluate the BCoSDD framework with two datasets of customers' locations, i.e., the Solomon benchmark suites~\cite{ref_solomon} and a real dataset from logistics industry in Singapore. The Solomon benchmark file {\ttfamily C101} is synthesized to have four depots and $60$ customers, i.e., $|c| = 60$. Alternatively, $100$ customers, i.e., $|c| = 100$, are considered for the real dataset experiments. The locations of depots and customers in the two datasets are shown in Figure~\ref{fig_map_solomon}. The number of customers is shared equally among the shippers. We set $C_1= \{c_1,c_5,\dots,c_{c'-3}\}$ to be the customers of shipper $p_1$, $C_2= \{c_2,c_6,\dots,c_{c'-2}\}$ to be the customers of shipper $p_2$, $C_3= \{c_3,c_7,\dots,c_{c'-1}\}$ to be the customers of shipper $p_3$, and $C_4= \{c_4,c_8,\dots,c_{c'}\}$ to be the customers of shipper $p_4$. 

We implement the optimization model of package assignment as a linear programming in GAMS~\cite{ref_gams} and, it is solved by a commercial solver CPLEX~\cite{ref_gams}. 


\subsection{Analysis of Cooperation Management}
\label{sec_analysis}

\begin{figure*}
\begin{subfigure}[t]{0.5\textwidth}
\centering
\includegraphics[width=\textwidth]{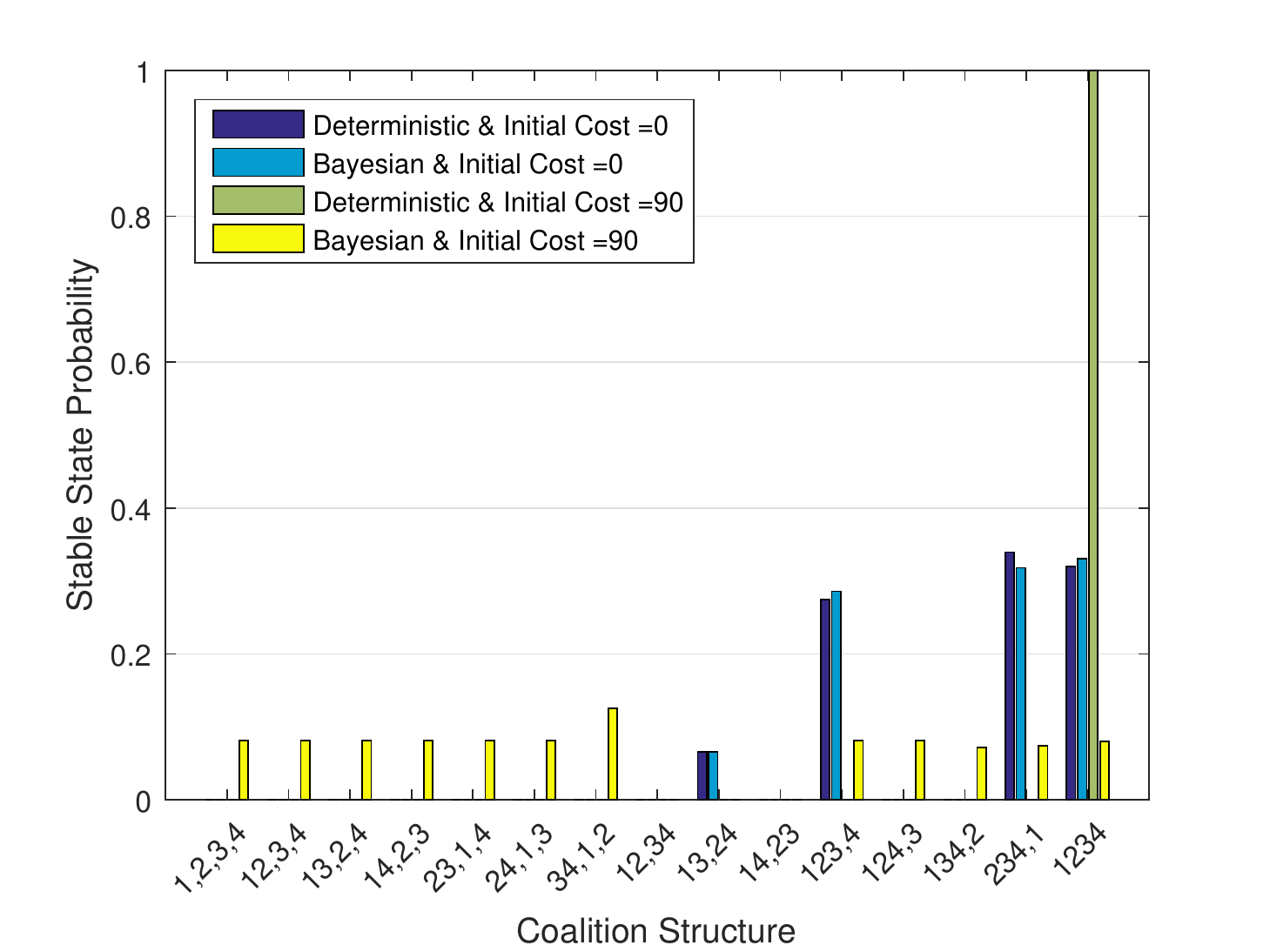}
\caption{Markov Transition Process}
\label{fig_ana_markov}
\end{subfigure}
\begin{subfigure}[t]{0.5\textwidth}
\centering
\includegraphics[width=\textwidth]{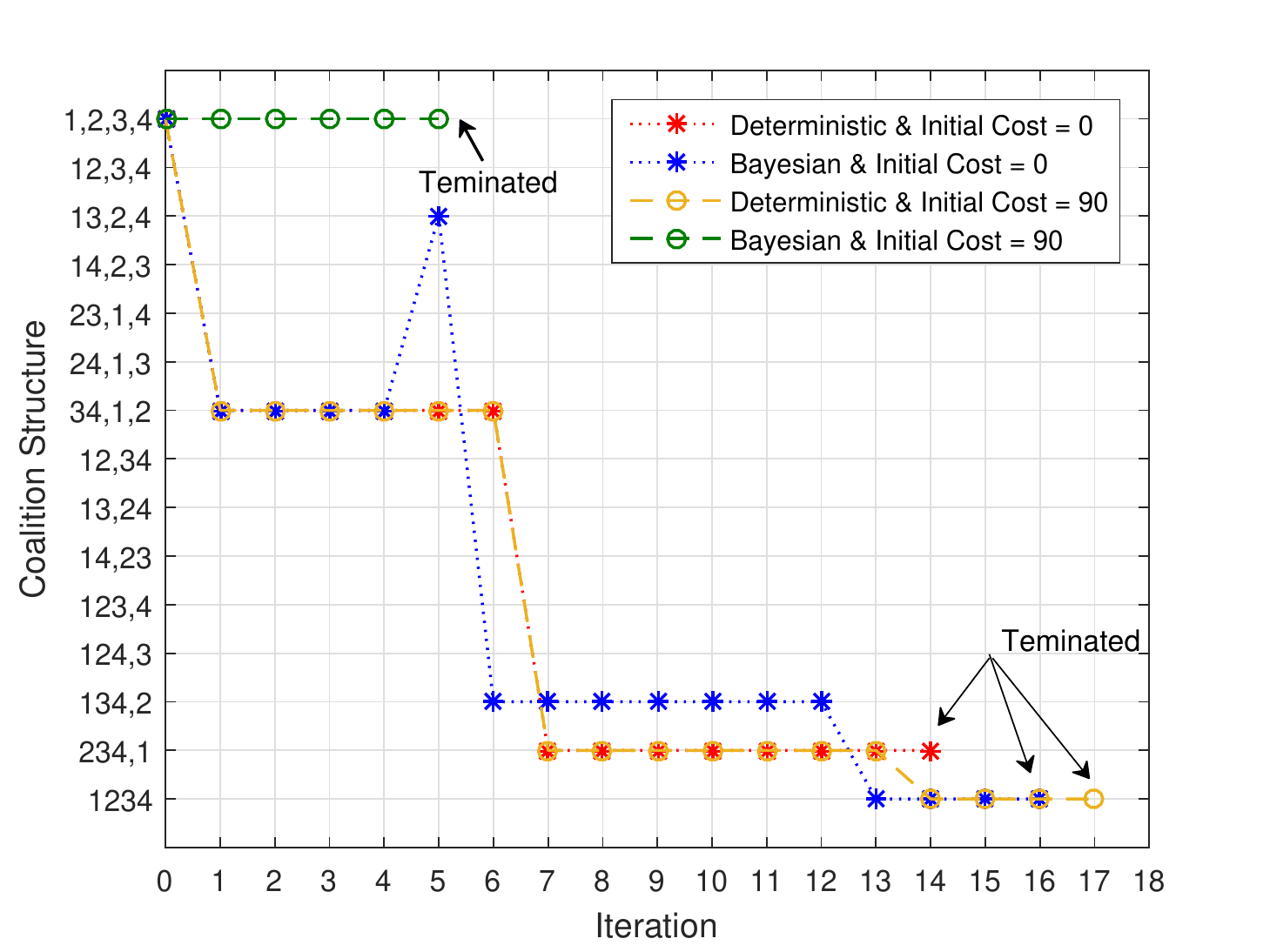}
\caption{Merge and Split Algorithm}
\label{fig_ana_merge}
\end{subfigure}
\caption{Analysis of the stable coalition formation}
\end{figure*}

The initial cost of drones is set as $C^{(i)}_d= S\$0$ and $C^{(i)}_d=S\$90$, where $C^{(i)}_d= S\$0$ is when a shipper has its own drone and does not need to pay for the initial cost. We consider the traditional approach which is referred to as deterministic approach, and the Bayesian approach, i.e., the BCoSDD framework. The deterministic approach considers the case that all shippers are the good cooperative shippers, and they always deliver all packages successfully. However, the Bayesian approach considers the case that all shippers believe that other cooperative shippers may fail to deliver packages with probability $P(b_{pq} = {\mathrm{T^B}}) = 1- P(b_{pq} = {\mathrm{T^G}})=0.1$ for all $p,q \in \mathcal{P}$. Table~\ref{table_shap} shows the individual delivery cost for all the coalition structures, i.e., $\Phi_1$ to $\Phi_{15}$. The stationary probabilities obtained from the Markov model are presented in Figure~\ref{fig_ana_markov}. The selected coalition structures of the merge and split algorithm are shown in Figure~\ref{fig_ana_merge}, where the last selection is the stable coalition structure. 

When we set $C^{(i)}_d = S\$0$, and there are three stable coalition structures in the deterministic and Bayesian approaches, i.e., $\Phi_{11}$, $\Phi_{14}$, and $\Phi_{15}$. The coalition structures $\Phi_{14}$ and $\Phi_{15}$ have the highest probabilities in the deterministic approach and the Bayesian approach, respectively. Identically, the merge and split algorithm achieves $\Phi_{14}$ and $\Phi_{15}$ as the stable coalition structures for the deterministic approach and the Bayesian approach, respectively. Since both shippers $p_1$ and $p_4$ want to cooperate with shipper $p_3$, and shipper $p_3$ can choose to cooperate such that its cost is minimized. As a result, shipper $p_3$ cooperates with shipper $p_4$, and the resulting coalition structure is $\Phi_{7}$ for the deterministic approach. By contrast, shipper $p_3$ cooperates with shipper $p_1$, and the resulting coalition structure is $\Phi_{3}$ for the Bayesian approach. For the deterministic approach, shippers $p_3$ and $p_4$ allow shipper $p_2$ to join the coalition in the seventh iteration, and the corresponding coalition structure is $\Phi_{14}$ to lower the cost. The stable coalition structure is $\Phi_{14}$ as shipper $p_4$ does not allow shipper $p_1$ to join the coalition because the shipper $p_4$ will pay a higher cost if coalition structure $\Phi_{15}$ is reached. On the other hand, coalition structure $\Phi_{15}$ is stable for the Bayesian approach because all the shippers achieve the lowest cost when this coalition structure is reached. 

When we set $C^{(i)}_d = S\$90$, $\Phi_{15}$ is the stable coalition structure from both the Markov model and the merge and split algorithm for the deterministic approach. However, there is no stable coalition structure for the Bayesian approach since all the stationary probabilities are less than $0.13$ and the merge and split algorithm terminates with the initial coalition structure. As such, the shippers should not cooperate with each other as the initial cost $C^{(i)}_d = S\$90$ is high, and hence the cooperation results in higher cost than outsourcing the package delivery to the carrier.

We have shown that the merge and split algorithm can achieve the same solution as that from the Markov model. While the Markov model requires the Shapley values of all the coalition structures before being solved, the merge and split algorithm needs only the Shapley values of the neighboring coalition structures when changing the selected coalition structure. For example, the merge and split algorithm does not require the Shapley values of coalition structures $\Phi_{11}$ and $\Phi_{12}$ in the experiment for the deterministic approach with $C^{(i)}_d = S\$0$. 


\begin{table*}
\caption{Shapley value and coalition structures}
\label{table_shap}
\scriptsize
\begin{tabular}{|p{13.37em}|p{1.8em}p{1.8em}p{1.8em}p{2.5em}|p{1.8em}p{1.8em}p{1.8em}p{2.5em}|p{1.8em}p{1.8em}p{1.8em}p{2.5em}|p{1.8em}p{1.8em}p{1.8em}p{2.5em}|}
\hline 
\multicolumn{1}{|c}{ }& \multicolumn{4}{|c|}{Deterministic (Initial Cost = 0)}& \multicolumn{4}{|c|}{Bayesian (Initial Cost = 0)} & \multicolumn{4}{|c|}{Deterministic (Initial Cost = 90)}& \multicolumn{4}{|c|}{Bayesian (Initial Cost = 90)} \\\hline 
Coaltition Structure & $P_1$ & $P_2$ & $P_3$ & $P_4$ & $P_1$ & $P_2$ & $P_3$ & $P_4$& $P_1$ & $P_2$ & $P_3$ & $P_4$ & $P_1$ & $P_2$ & $P_3$ & $P_4$\\
\hline
$\Phi_1$= \{\{$p_1$\},\{$p_2$\},\{$p_3$\},\{$p_4$\}\} &174.74 &211.01 &186.13 &154.33	&174.74 &211.01 &186.13 &154.33
&240.00 &240.00 &240.00 &240.00	 &\textbf{240.00} &\textbf{240.00} &\textbf{240.00} &\textbf{240.00}\\
$\Phi_2$= \{\{$p_1$,$p_2$\},\{$p_3$\},\{$p_4$\}\} 	&174.75 &211.02 &186.13 &154.33	&174.75 &211.02 &186.13 &154.33
&240.00 &240.00 &240.00 &240.00	 &240.00 &240.00 &240.00 &240.00\\
$\Phi_3$= \{\{$p_1$,$p_3$\},\{$p_2$\},\{$p_4$\}\} 	&158.05 &211.01 &169.44 &154.33	 &166.05 &211.01 &175.84 &154.33
&240.00 &240.00 &240.00 &240.00	 &240.00 &240.00 &240.00 &240.00\\
$\Phi_4$= \{\{$p_1$,$p_4$\},\{$p_3$\},\{$p_3$\}\} &174.74 &211.01 &186.13 &154.33	 &174.74 &211.01 &186.13 &154.33
&240.00 &240.00 &240.00 &240.00	 &240.00 &240.00 &240.00 &240.00\\	
$\Phi_5$= \{\{$p_2$,$p_3$\},\{$p_1$\},\{$p_4$\}\} 	&174.74 &211.01 &186.13 &154.33	 &174.74 &211.01 &186.13 &154.33
&240.00 &240.00 &240.00 &240.00	 &240.00 &240.00 &240.00 &240.00\\
$\Phi_6$= \{\{$p_2$,$p_4$\},\{$p_1$\},\{$p_3$\}\} 	&174.74 &202.38 &186.13 &145.70	 &174.74 &211.98 &186.13 &148.90
&240.00 &240.00 &240.00 &240.00	 &240.00 &240.00 &240.00 &240.00\\
$\Phi_7$= \{\{$p_3$,$p_4$\},\{$p_1$\},\{$p_2$\}\} 	&174.74 &211.01 &165.49 &133.69	 &174.74 &211.01 &178.29 &140.09
 &240.00 &240.00 &236.87 &236.87	 &240.00 &240.00 &251.27 &236.87\\
$\Phi_8$= \{\{$p_1$,$p_2$\},\{$p_3$,$p_4$\}\} 		&174.75 &211.02 &165.49 &133.69	 &174.75 &211.02 &178.29 &140.09
 &240.00 &240.00 &236.87 &236.87	 &240.00 &240.00 &251.27 &236.87\\
$\Phi_9$= \{\{$p_1$,$p_3$\},\{$p_2$,$p_4$\}\} 		&158.05 &202.38 &169.44 &145.70	 &166.05 &211.98 &175.84 &148.90
&240.00 &240.00 &240.00 &240.00	 &240.00 &240.00 &240.00 &240.00\\
$\Phi_{10}$= \{\{$p_1$,$p_4$\},\{$p_2$,$p_3$\}\} 	&174.74 &211.01 &186.13 &154.33	 &174.74 &211.01 &186.13 &154.33
&240.00 &240.00 &240.00 &240.00	 &240.00 &240.00 &240.00 &240.00\\
$\Phi_{11}$= \{\{$p_1$,$p_2$,$p_3$\},\{$p_4$\}\} 	&137.19 &190.15 &148.57 &154.33	 &148.39 &198.15 &158.17 &154.33
&240.00 &240.00 &240.00 &240.00	 &240.00 &240.00 &240.00 &240.00\\
$\Phi_{12}$= \{\{$p_1$,$p_2$,$p_4$\},\{$p_3$\}\} 	&167.69 &195.33 &186.13 &138.65	 &174.09 &204.93 &186.13 &148.24
&240.00 &240.00 &240.00 &240.00	 &240.00 &240.00 &240.00 &240.00\\
$\Phi_{13}$= \{\{$p_1$,$p_3$,$p_4$\},\{$p_2$\}\} 	&152.77 &211.01 &143.52 &128.41	 &163.97 &211.01 &159.52 &141.21
 &229.97 &240.00 &226.84 &226.84	 &244.37 &240.00 &239.64 &233.24\\
$\Phi_{14}$= \{\{$p_2$,$p_3$,$p_4$\},\{$p_1$\}\} 	&\textbf{174.74} &\textbf{184.64} &\textbf{147.75} &\textbf{107.31} &174.74 &199.04 &162.15 &116.91	& 240.00 &234.65 &231.52 &231.52	 &240.00 &249.05 &242.71 &237.92\\
$\Phi_{15}^*$ = \{\{$p_1$,$p_2$,$p_3$,$p_4$\}\} 	&138.61 &170.48 &118.67 &108.74	 &\textbf{153.01} &\textbf{184.88} &\textbf{136.27} &\textbf{124.74}	 & \textbf{227.20} &\textbf{231.88} &\textbf{218.71} &\textbf{218.71} &241.60 &246.28 &236.31 &234.71\\
\hline
\end{tabular}
\end{table*}

\subsection{Impact of the Optimization}
\label{sec_opt}

\begin{figure*}
\scriptsize
\centering
$\begin{array}{ccc}
\includegraphics[width=0.35\textwidth]{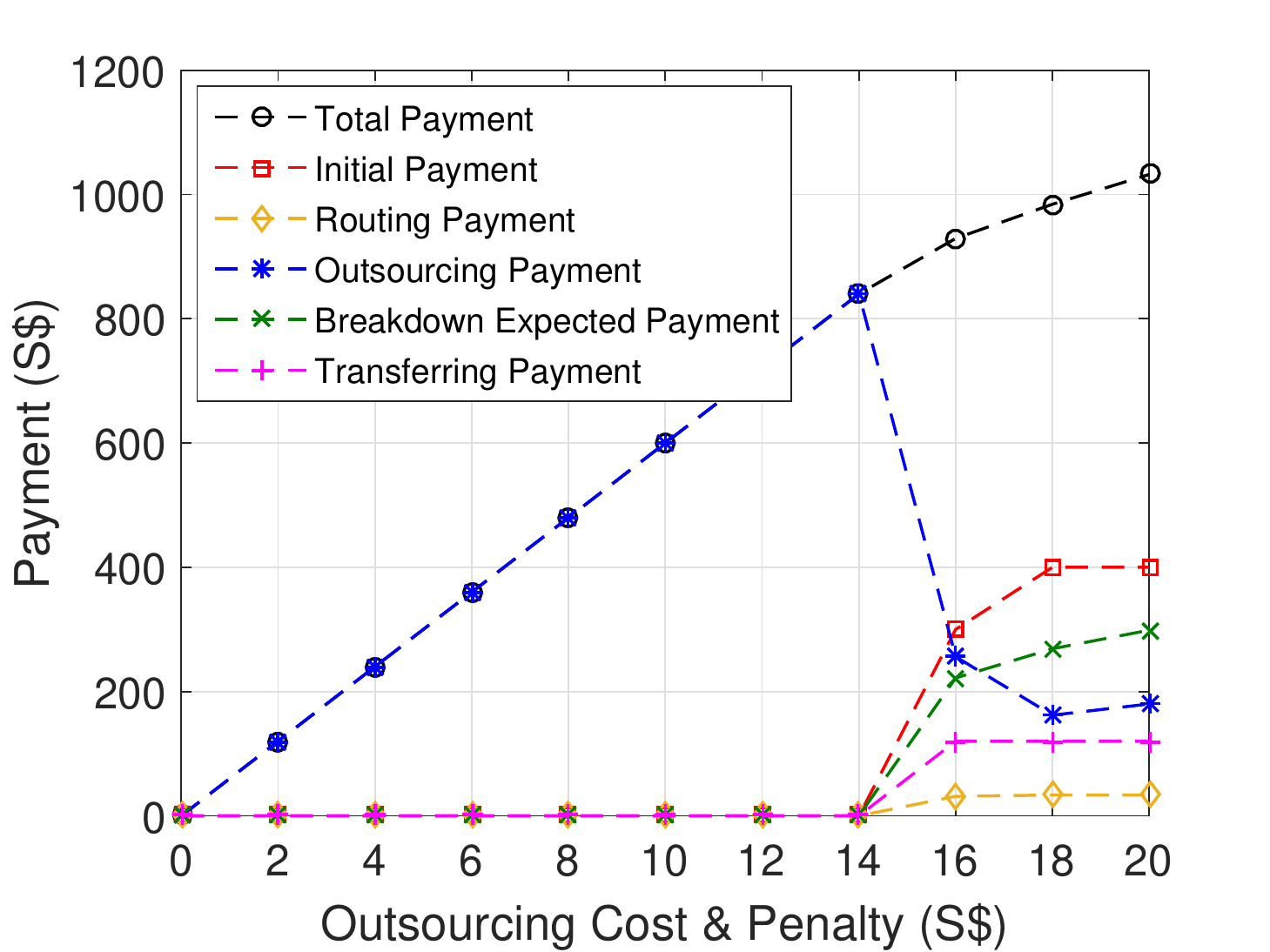}
&\hspace{-2em}\includegraphics[width=0.35\textwidth]{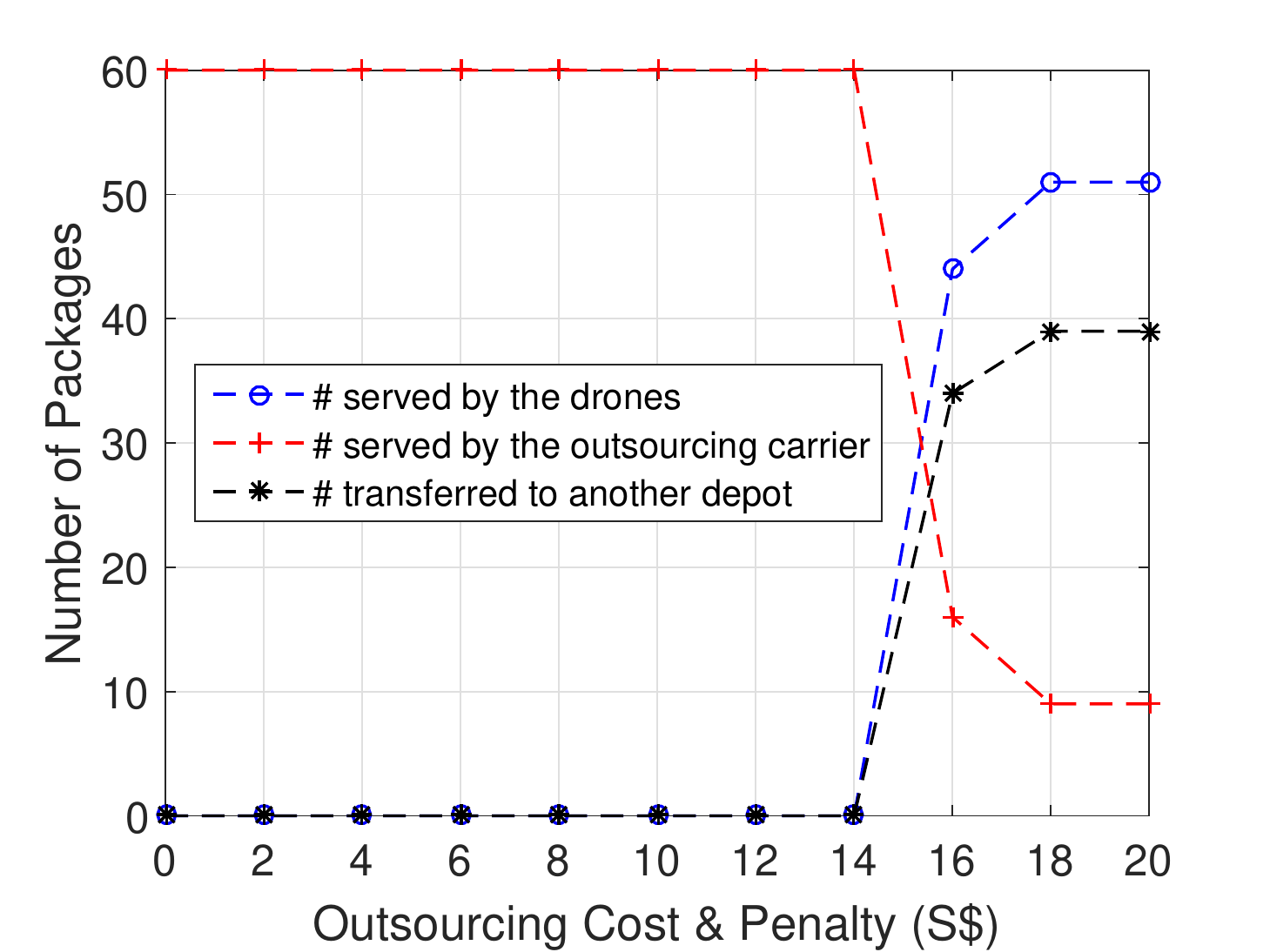}
&\hspace{-2em}\includegraphics[width=0.35\textwidth]{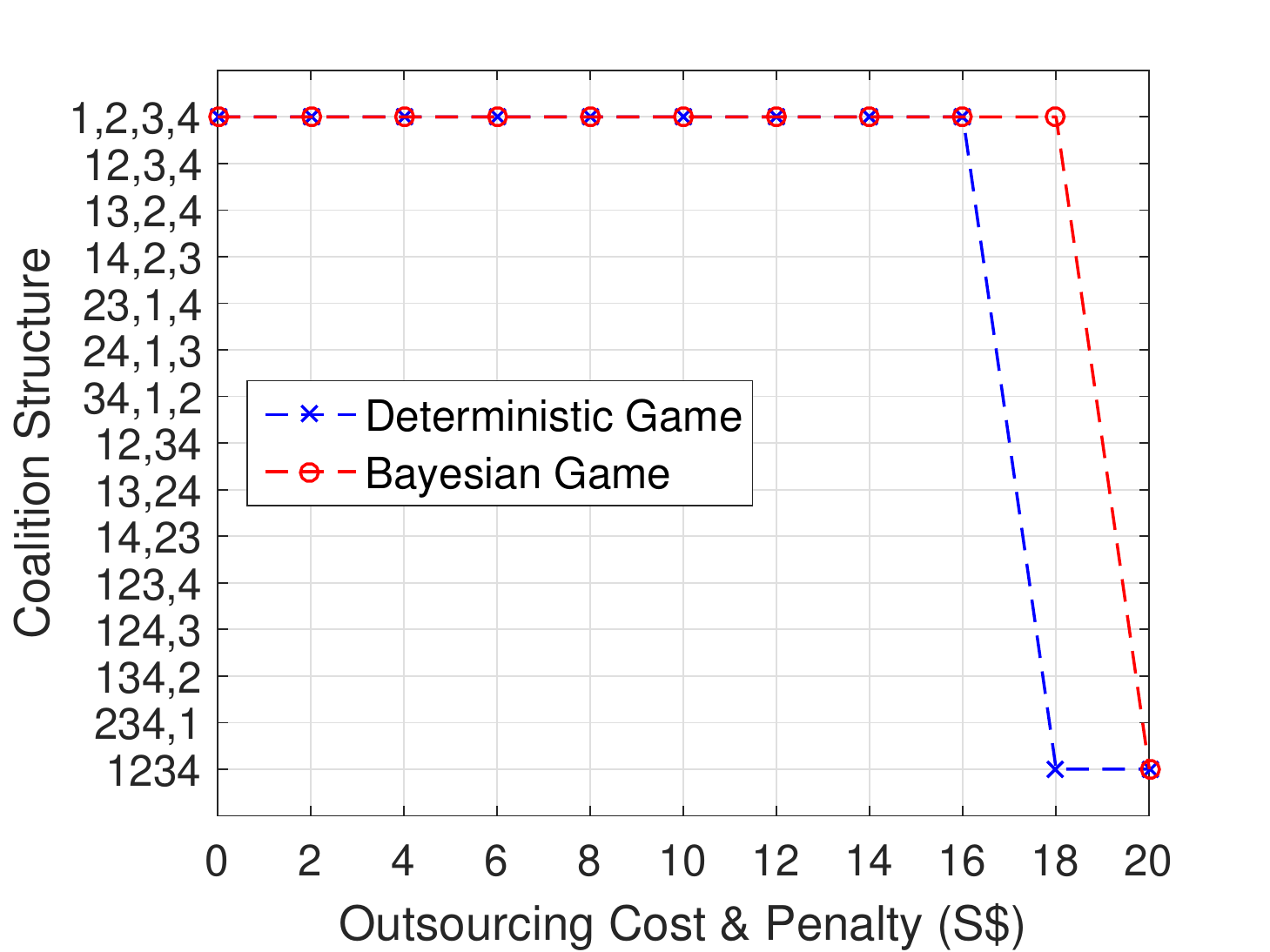}\\
(a)-Penalty&(b)-Penalty&(c)-Penalty\\
\includegraphics[width=0.35\textwidth]{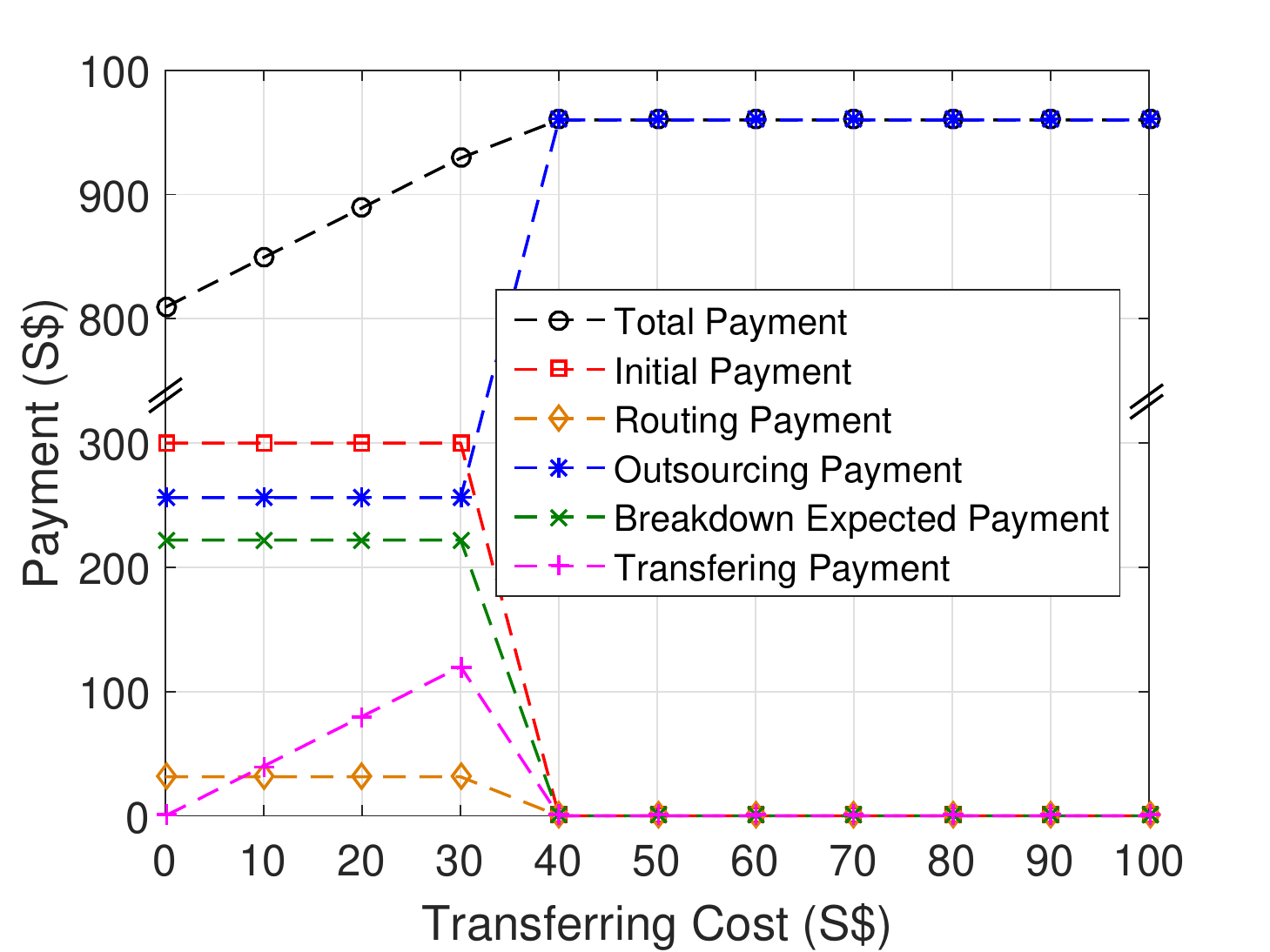}
&\hspace{-2em}\includegraphics[width=0.35\textwidth]{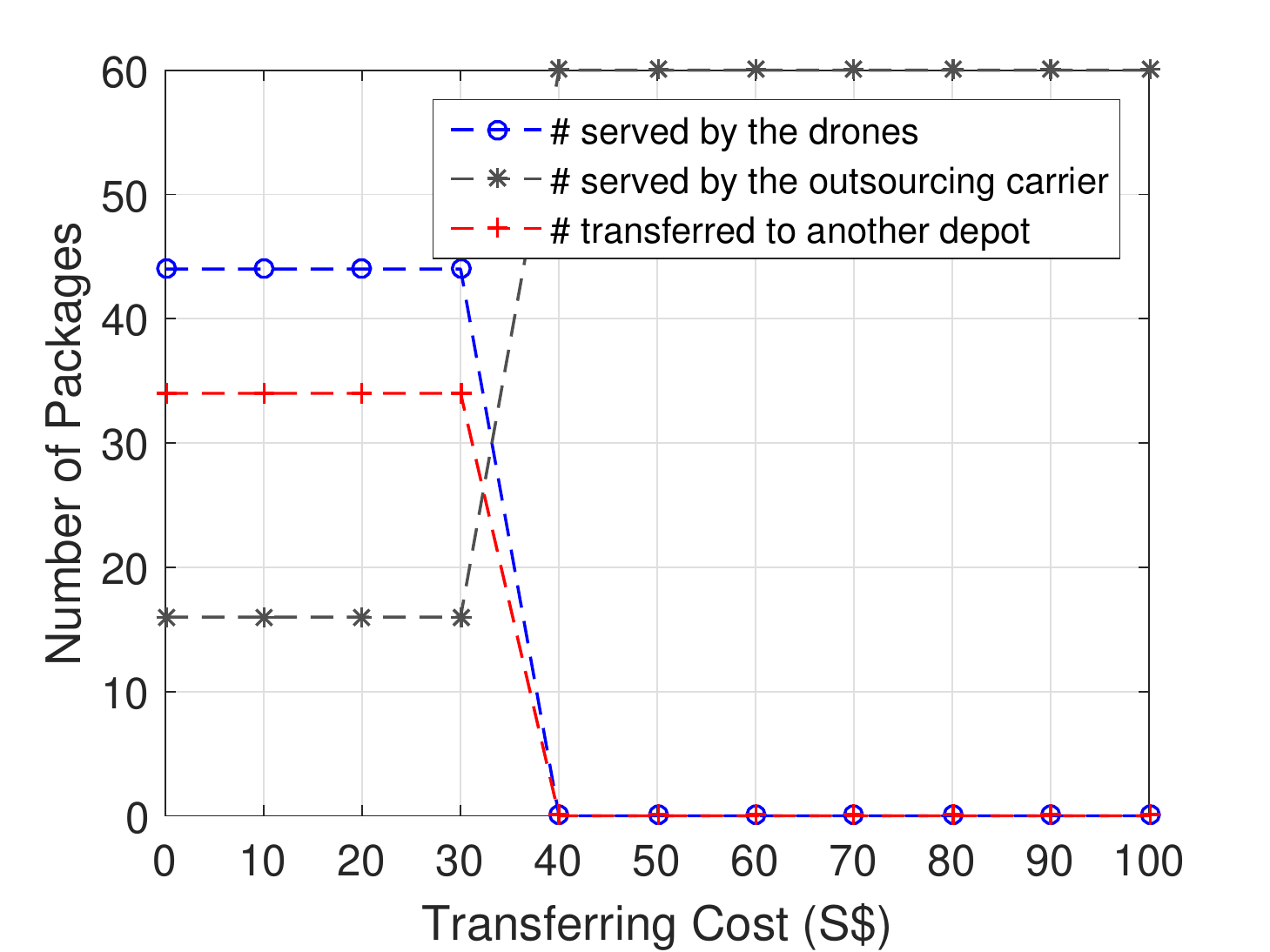}
&\hspace{-2em}\includegraphics[width=0.35\textwidth]{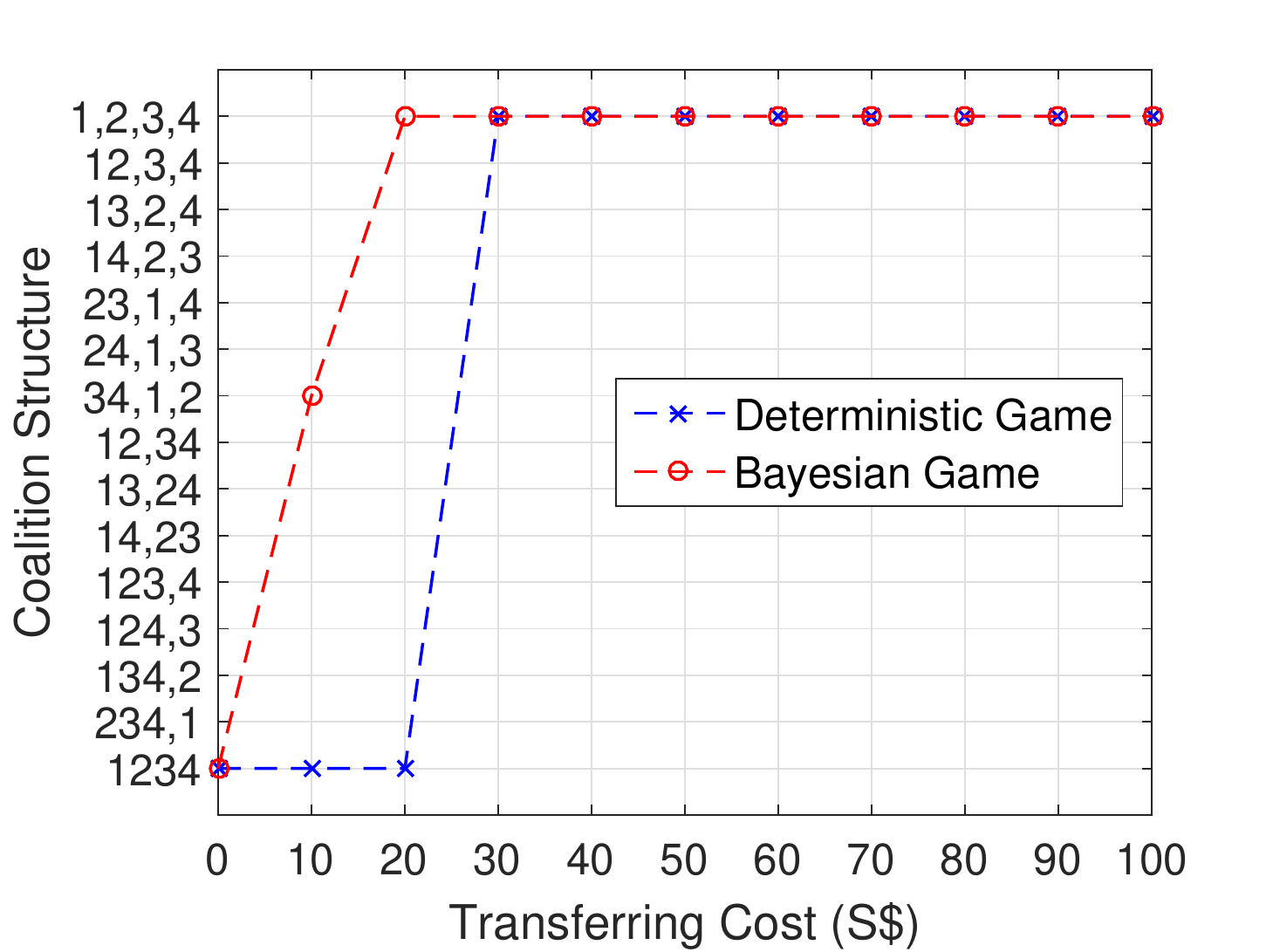}\\
(d)-Transfer&(e)-Transfer&(f)-Transfer\\
\includegraphics[width=0.35\textwidth]{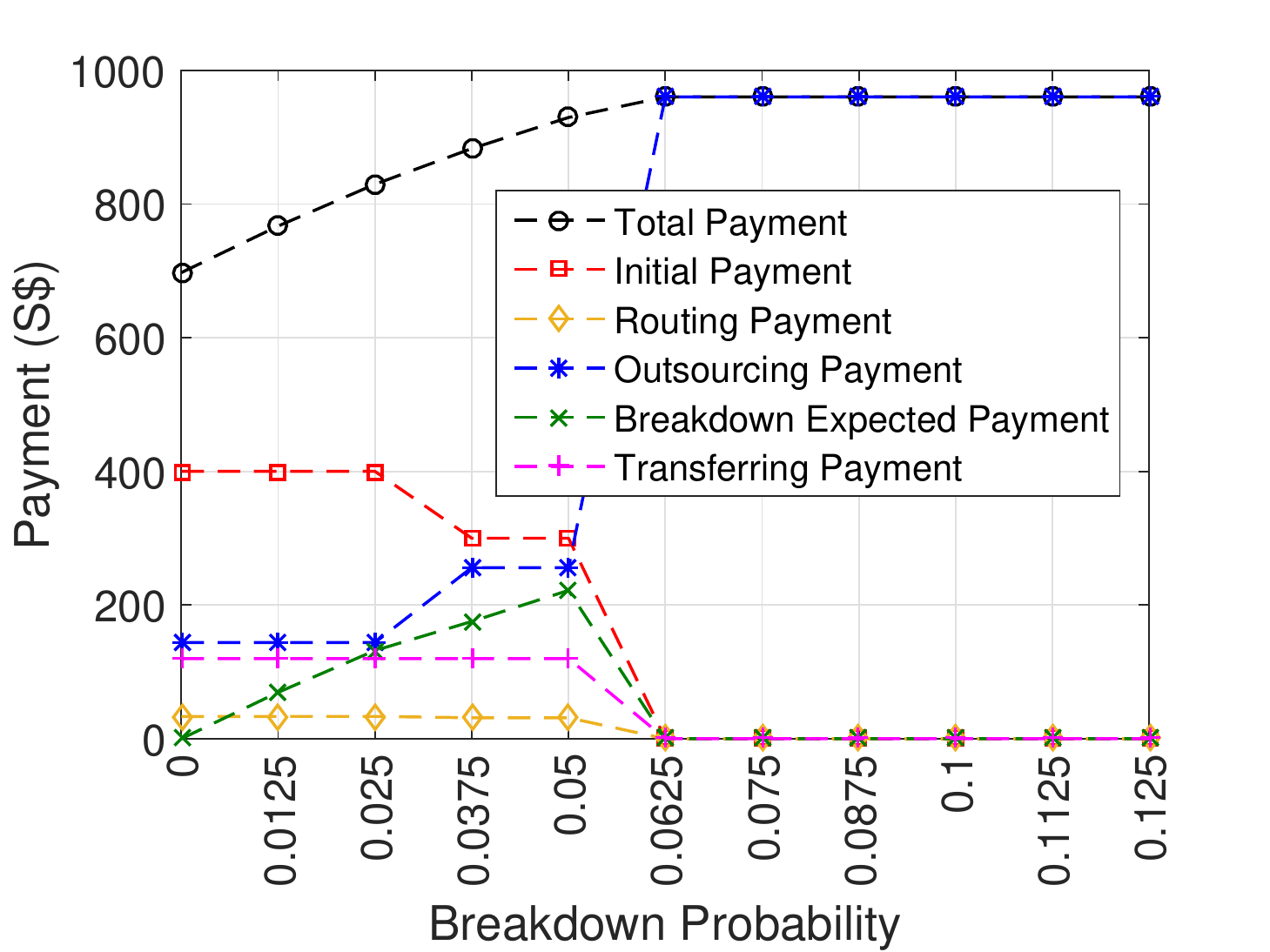}
&\hspace{-2em}\includegraphics[width=0.35\textwidth]{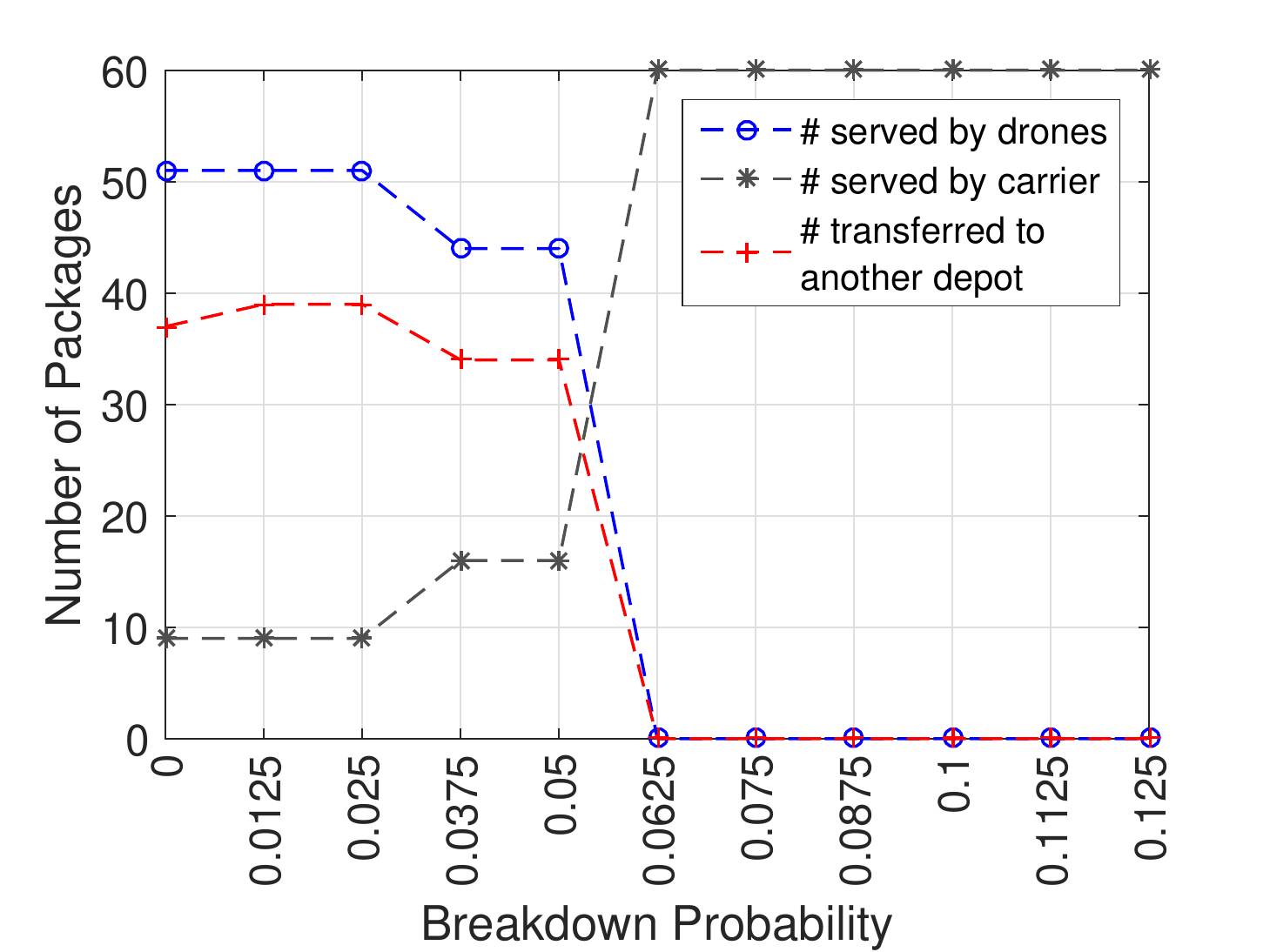}
&\hspace{-2em}\includegraphics[width=0.35\textwidth]{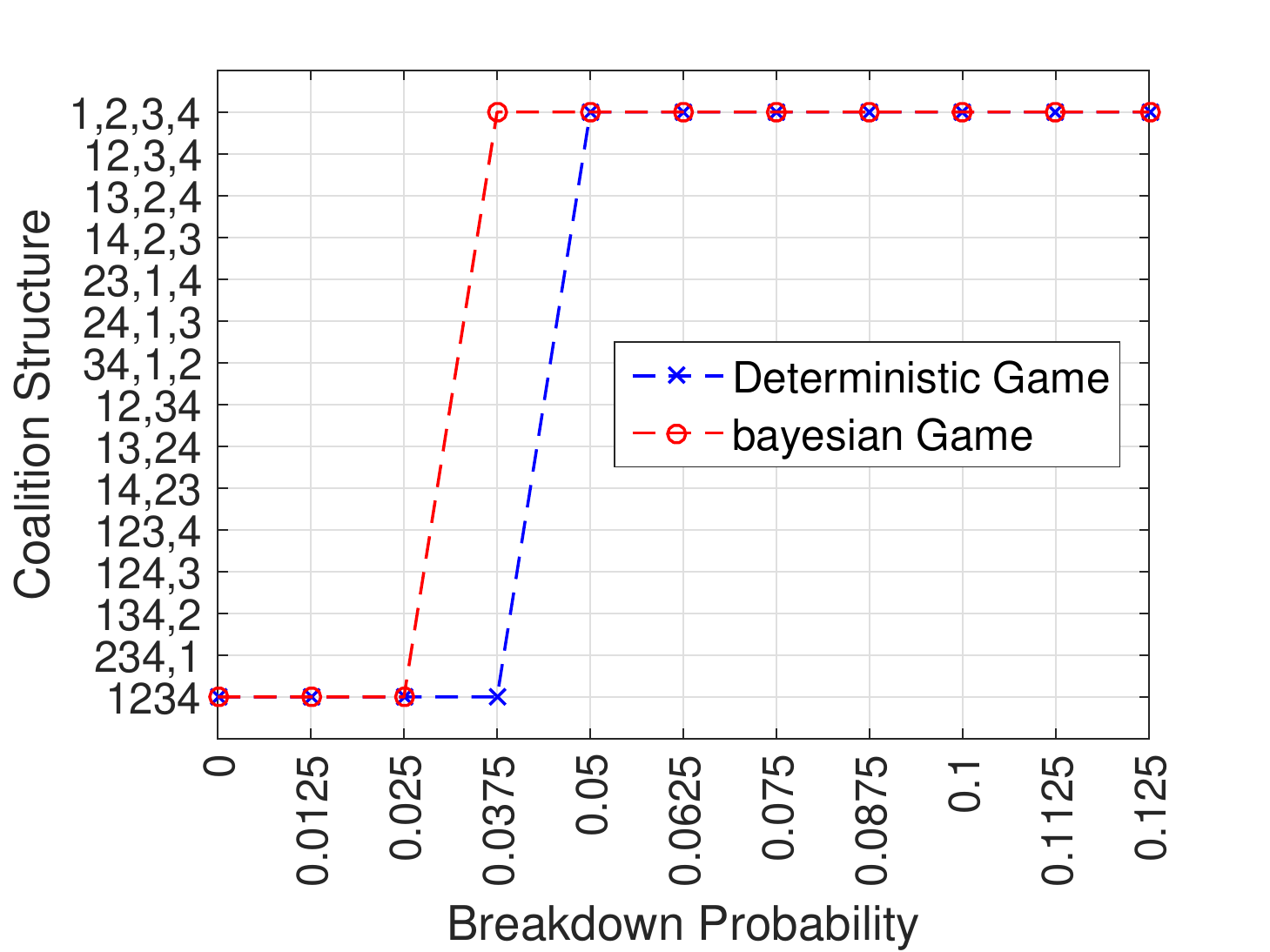}\\
(g)-Breakdown&(h)-Breakdown&(i)-Breakdown\\
\includegraphics[width=0.35\textwidth]{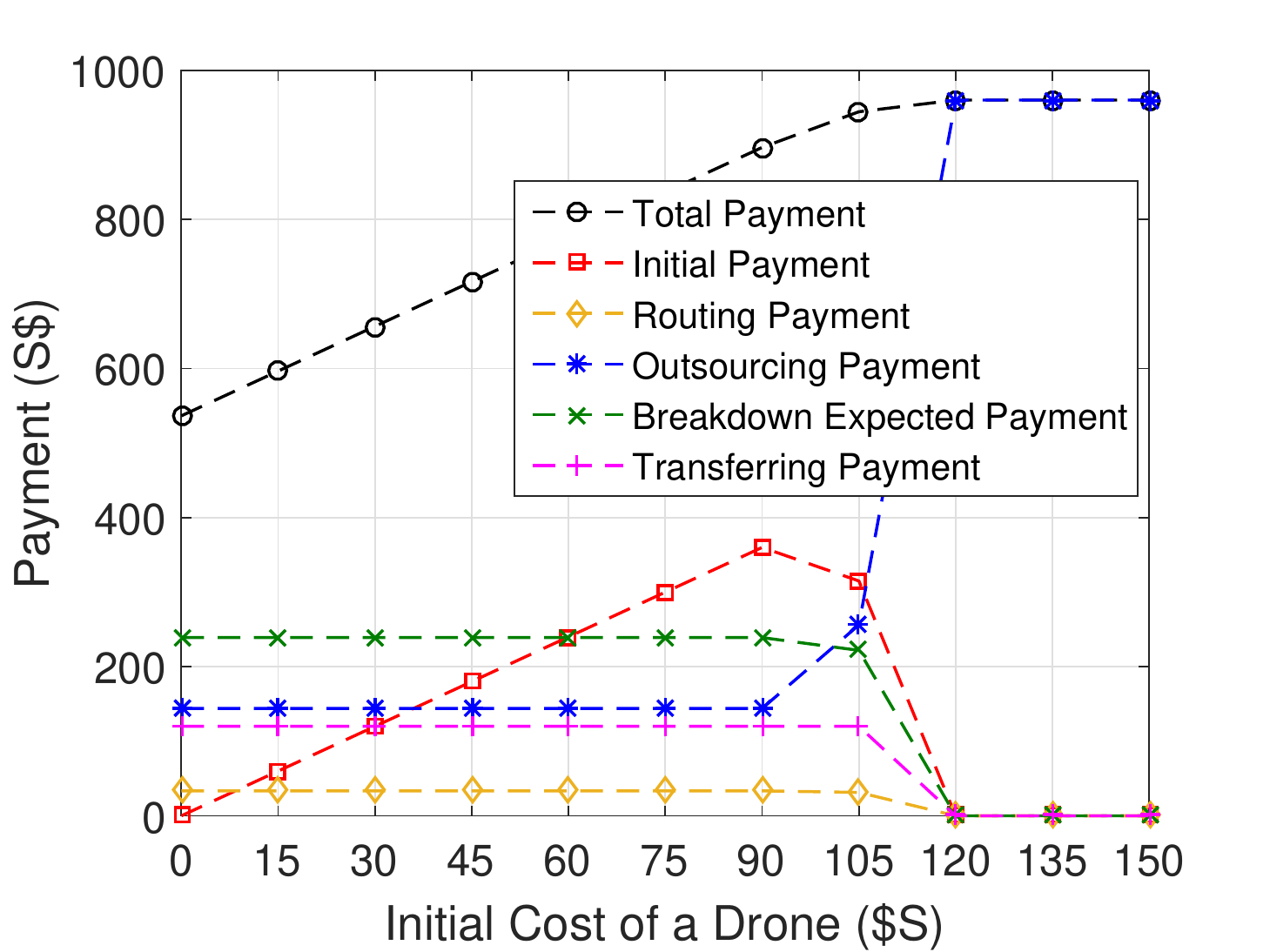}
&\hspace{-2em}\includegraphics[width=0.35\textwidth]{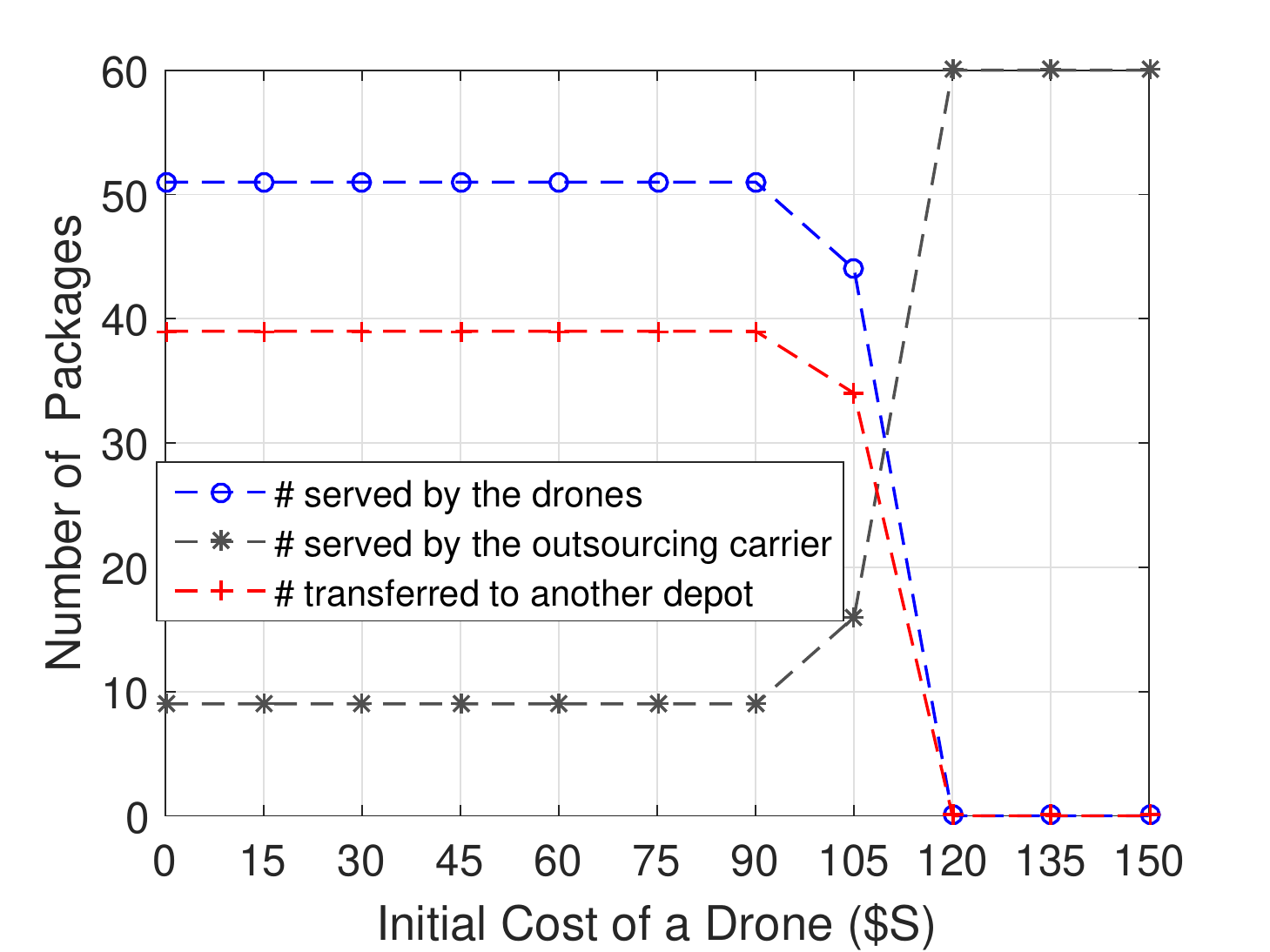}
&\hspace{-2em}\includegraphics[width=0.35\textwidth]{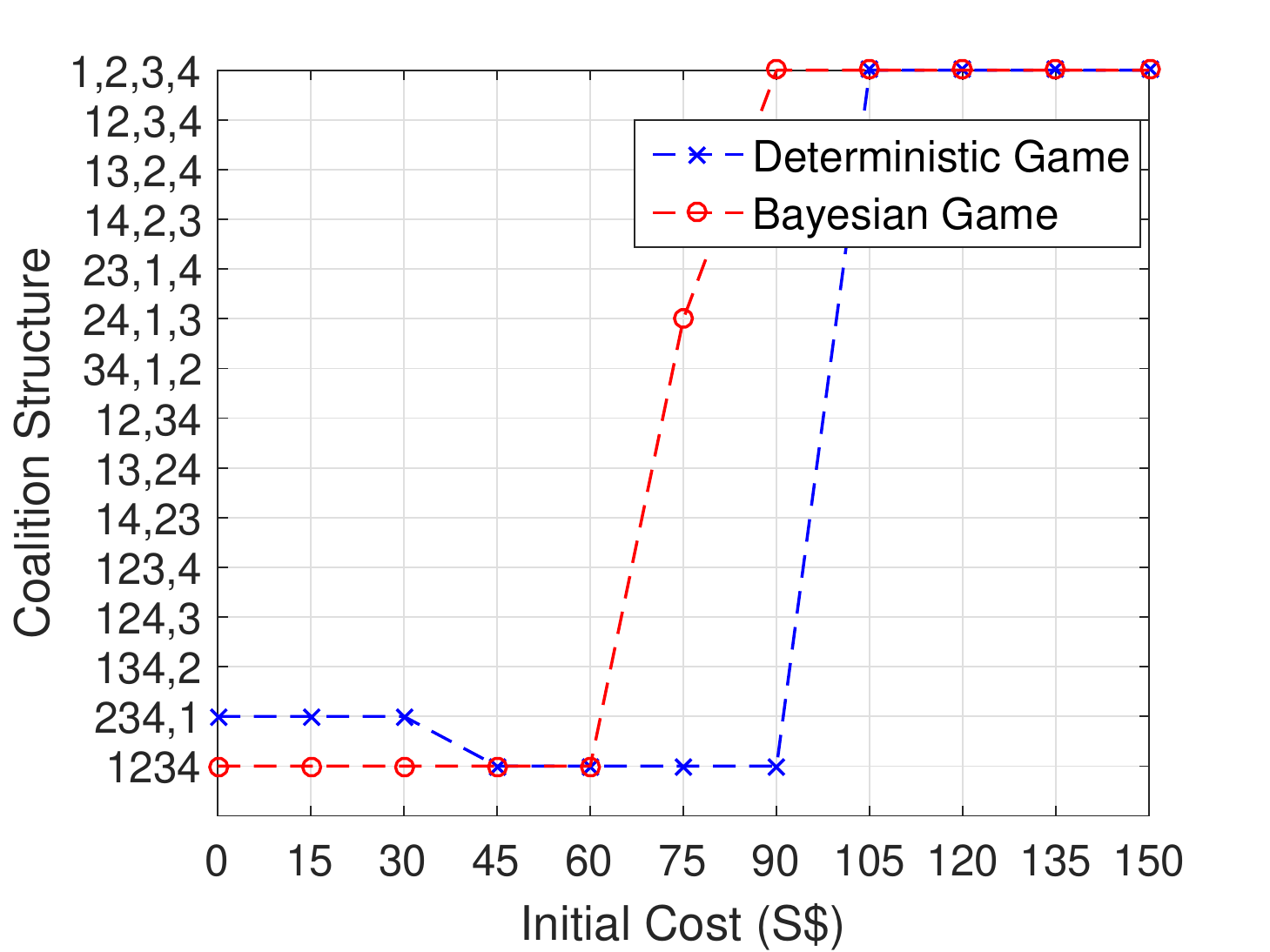}\\
(j)-InitialCost&(k)-InitialCost&(l)-InitialCost\
\end{array}$
\caption{Impact of Optimization}
\label{fig_opt}
\end{figure*}

To understand the results in the cooperation management more deeply, we thus evaluate only the package assignment of the BCoSDD framework. We conduct four experiments including (i) varying the penalty $C^{(p)}_i$ and the outsourcing cost $C^{(c)}_i$, (ii) varying the transferring cost $C^{(t)}_p$, (iii) varying the breakdown probability $\mathbb{P}_d$, and (iv) varying the initial cost $C^{(i)}_d$. Figures~\ref{fig_opt}(a), (e), (g), and (j) show the costs of the coalition, when all the shippers cooperate, i.e., the coalition structure $\Phi_{15}$. Similarly, Figures~\ref{fig_opt}(b), (e), (h), and (k) show the numbers of customers which are served by the drones, served by the outsourcing carrier, and whose packages are transferred to another depot under the coalition structure $\Phi_{15}$. Figures~\ref{fig_opt}(c), (f), (i), and (l) show the impact of the parameters to the stable coalitions.

When the penalty and the outsourcing cost are set such that $C^{(p)} = C^{(c)}_i$ and are varied, the shippers outsource all packages to the carrier when the outsourcing cost is cheap, i.e. $C^{(c)}_i \leq S\$14$. As a result, the shippers do not cooperate. If all the shippers are cooperate ($\Phi_{15}$), three drones are used to serve $44$ customers for $C^{(c)}_i = S\$16$, and four drones are used to serve $51$ customers for $S\$ 16\leq C^{(c)}_i \leq S\$20$. The remaining customers are served by the carrier. Note that the drone cannot serve $9$ customers because they are outside the serving area as indicated in Figure~\ref{fig_map_solomon}. 

All the shippers will cooperate when $C^{(c)}_i \geq S\$18$ for the deterministic approach and $C^{(c)}_i = S\$20$ in the Bayesian approach. However, the experiment of varying transferring cost ($C^{(t)}_p$) has a different result. Specifically, all the shippers cooperate ($\Phi_{15}$) when the transferring cost is cheap ($C^{(t)}_p = S\$0$), and they do not cooperate when the transferring cost is high. If all of the shippers cooperate, i.e., $\Phi_{15}$, three drones are used to serve $44$ customers, and $34$ packages are transferred when $C^{(t)}_p \leq S\$30$. Moreover, when all the shippers cooperate ($\Phi_{15}$), the experiments of varying the breakdown probability and varying the initial cost share a similar trend. In particular, four drones are used when the breakdown probability is small ($\mathbb{P}_d \leq 0.025$) and when the initial cost is cheap ($C^{(i)}_d \leq S\$90$). After the parameters increase to a certain point, i.e., $\mathbb{P}_d =0.0375$ and $C^{(i)}_d = S\$105$, the number of allocated drones decreases to three. After increasing the parameters further until the drone delivery cost is higher than the outsourcing cost, no drone is used when $\mathbb{P}_d \geq 0.0625$ and $C^{(i)}_d \geq S\$120$ for the experiment of varying breakdown probability and varying the initial cost, respectively.

Based on Figure~\ref{fig_opt}c, we can conclude that the stable coalition structure is affected by the system parameters. From the experiments, the deterministic approach tends to let the shippers join a coalition easier and leave the coalition harder than the Bayesian approach. Again, the deterministic approach considers the case that all shippers always deliver all packages successfully. For example, when the outsourcing cost is high, the deterministic approach lets the shippers form coalition structure $\Phi_{15}$ since $C^{(c)}_i = S\$18$ while the Bayesian approach lets them deliver individually and form the same coalition later at $C^{(c)}_i = S\$20$. In contrast, the Bayesian approach lets shippers $p_1$ and $p_2$ leave the coalition $\Phi_{15}$ immediately after the transferring cost increases to $C^{(t)}_p =S\$10$ while the deterministic approach lets all the shippers cooperate $\Phi_{15}$ until $C^{(t)}_p = S\$20$. Note that we have explained the impact of the initial cost in Section~\ref{sec_analysis}, especially the reason that the stable coalition structure is $\Phi_{14}$ when the initial cost is small, i.e., $C^{(i)} = S\$0$.

\subsection{Simulation}
\label{sec_sim}

\begin{figure}
\includegraphics[width=0.5\textwidth]{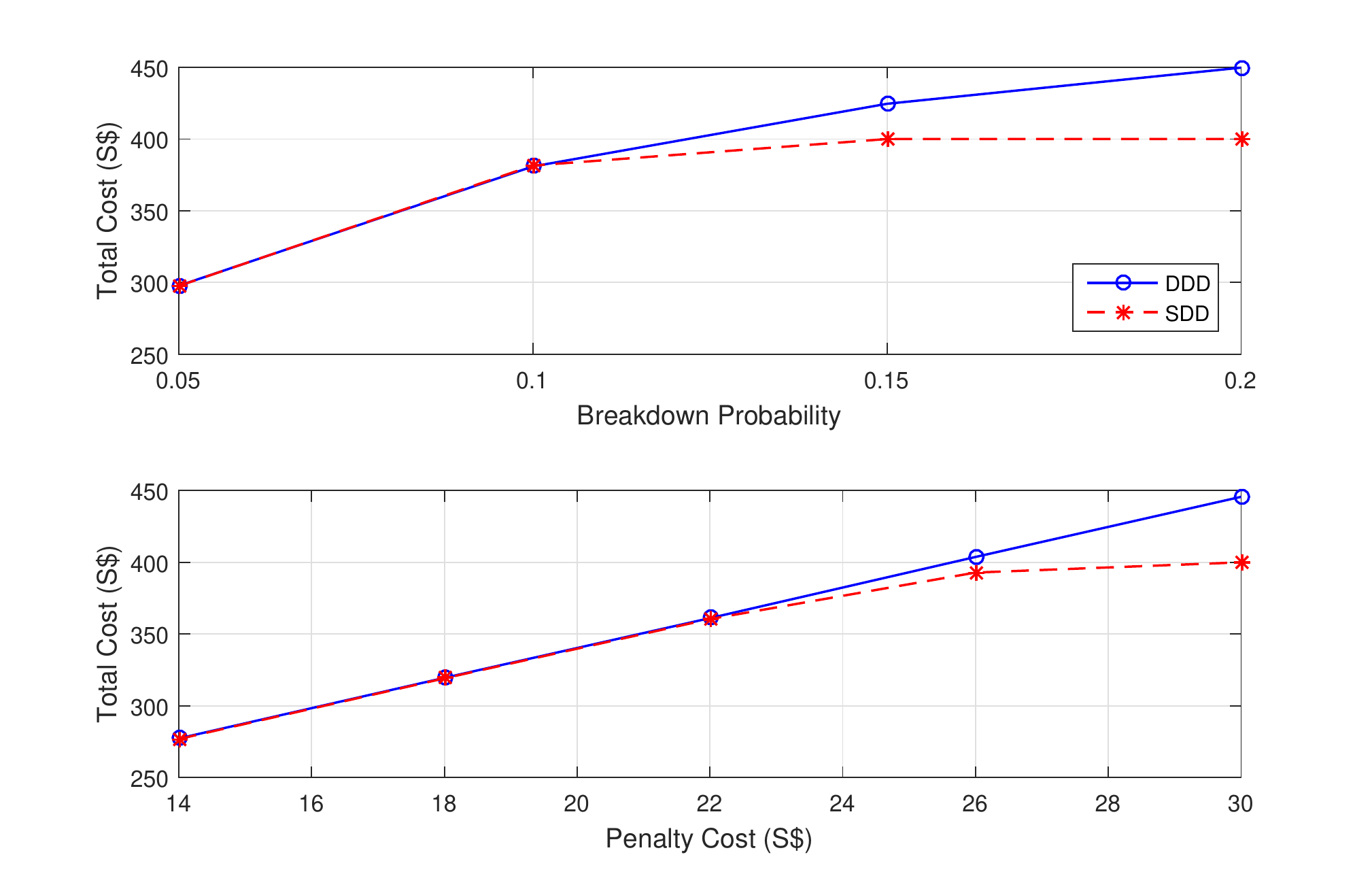}
\caption{Simulation results of deterministic (DDD) and stochastic (SDD) approaches of shipper $P_3$ with one drone. }
\label{fig_sim_opt}
\end{figure}

\begin{figure}
\includegraphics[width=0.5\textwidth]{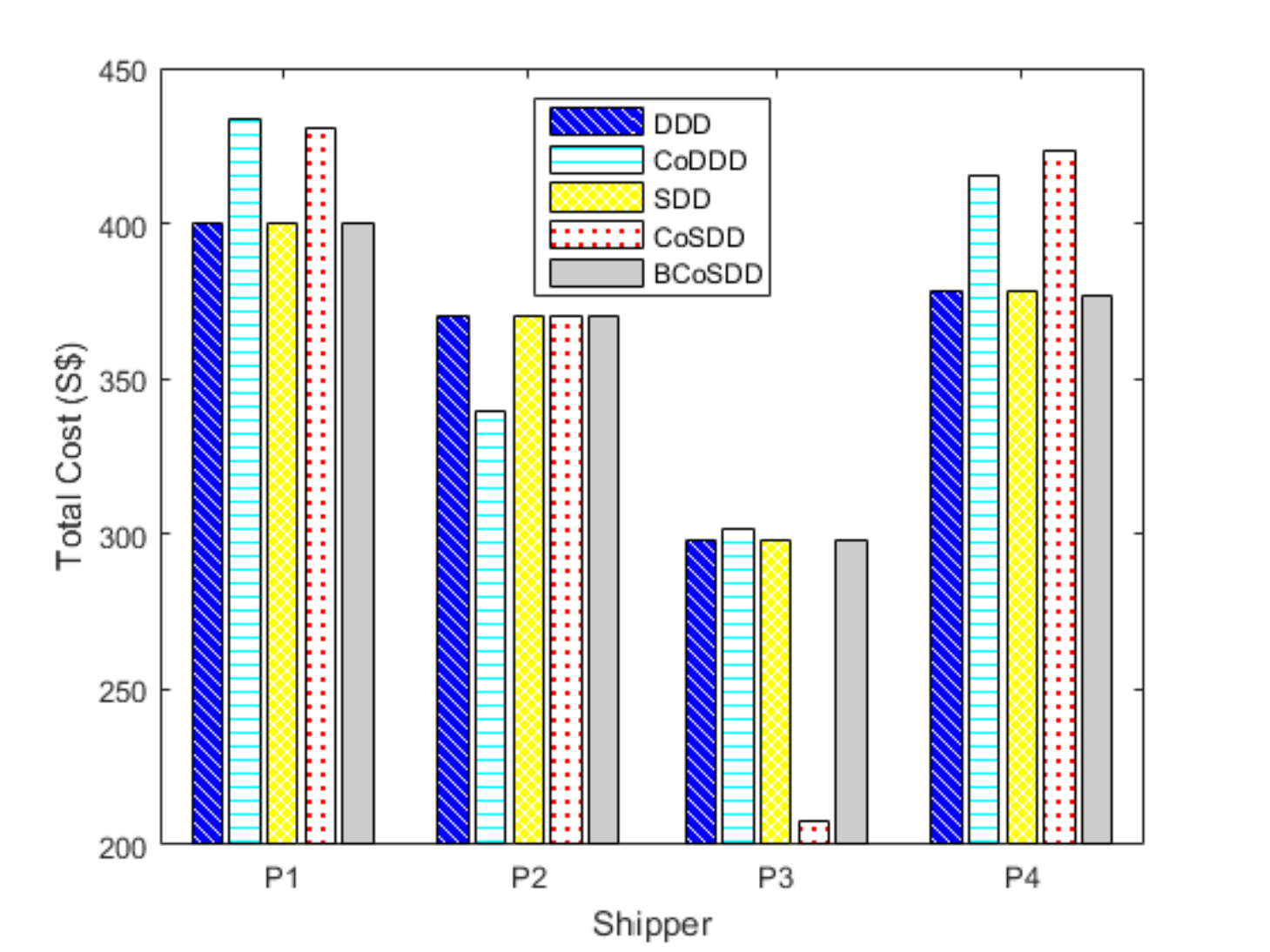}
\caption{The comparison of simulation results.}
\label{fig_sim_com}
\end{figure}

We also conduct the simulation experiments to validate the numerical results in which breakdown events are simulated.  
\begin{itemize}
	\item In the first experiment, we focus on the package assignment with two cases. In the first case, the package assignment does not consider the breakdown when obtaining the solution, and in the second case, it considers the breakdown. The first and second cases are referred to as the deterministic drone delivery (DDD) and the stochastic drone delivery (SDD), shown in Figure~\ref{fig_sim_opt}, respectively. 
	\item In the second experiment, we focus on the package assignment, cost sharing, and cooperation management. We compare our BCoSDD framework and other approaches in Figure~\ref{fig_sim_com}. In DDD and SDD approaches, none of the shippers cooperate. In CoDDD and CoSDD approaches, the shippers can cooperate without considering misbehavior. Nonetheless, in all the approaches, the cooperative shippers can misbehave.
\end{itemize}

We experiment the SDD approach with only one shipper, i.e., $p_3$. From Figure~\ref{fig_sim_opt}, the total cost achieved from the SDD approach is lower than or equal to that of the DDD approach. When the breakdown probability and penalty cost increase, the SDD approach tends to assign packages to the carrier more easily than the DDD approach because the DDD approach does not consider the breakdown. As such, the SDD approach can avoid the penalty cost better than the DDD approach. Therefore, the SDD approach achieves the lower total cost than that of the DDD approach when the breakdown probability is $0.15 \leq \mathbb{P}_d \leq 0.2$ and when the breakdown penalty is $S\$26 \leq C^{(p)} \leq S\$30$.

Furthermore, Figure~\ref{fig_sim_com} presents the individual cost that each shipper needs to pay when using different approaches. The BCoSDD approach achieves the stable coalition structure $\Phi_6$, which shippers $p_2$ and $p_4$ cooperate. However, the CoDDD and CoSDD approaches achieve the stable coalition structures $\Phi_{13}$ and $\Phi_{15}$. Based on the simulation results, the solution from the CoDDD approach favors shipper $p_2$, but the solution is not fair to the other shippers as they will pay a higher delivery cost than that when they does not cooperate. Similarly, the solution from the CoSDD approach favors shipper $p_3$. Apparently, the distributed cost for each shipper from the BCoSDD approach is always relational and the lowest cost compared with the DDD, CoDDD, SDD, and CoSDD approaches.

\subsection{Belief Update Mechanism}
\label{sec_update}

The belief update mechanism is used in the BCoSDD framework in multiple iterations. We evaluate the belief update mechanism of the BCoSDD framework by assuming that shippers $p_1$ and $p_2$ always deliver all packages when they cooperate, i.e., good cooperative shippers, while shippers $p_3$ and $p_4$ will misbehave with a probability of 0.25, i.e., they are the bad cooperative shippers for 25\% of the time. Therefore, the shipper whose packages are unsuccessfully delivered by shippers $p_3$ and $p_4$ needs to pay the penalty. All the shippers are initialized to believe that the other shippers will deliver all packages, i.e., $P(b_{pq} = {\mathrm{T^G}}) = 1$ and $P(b_{pq} = {\mathrm{T^B}}) = 0$. 

\begin{figure}
\begin{subfigure}[t]{0.5\textwidth}
\centering
\includegraphics[width=\textwidth]{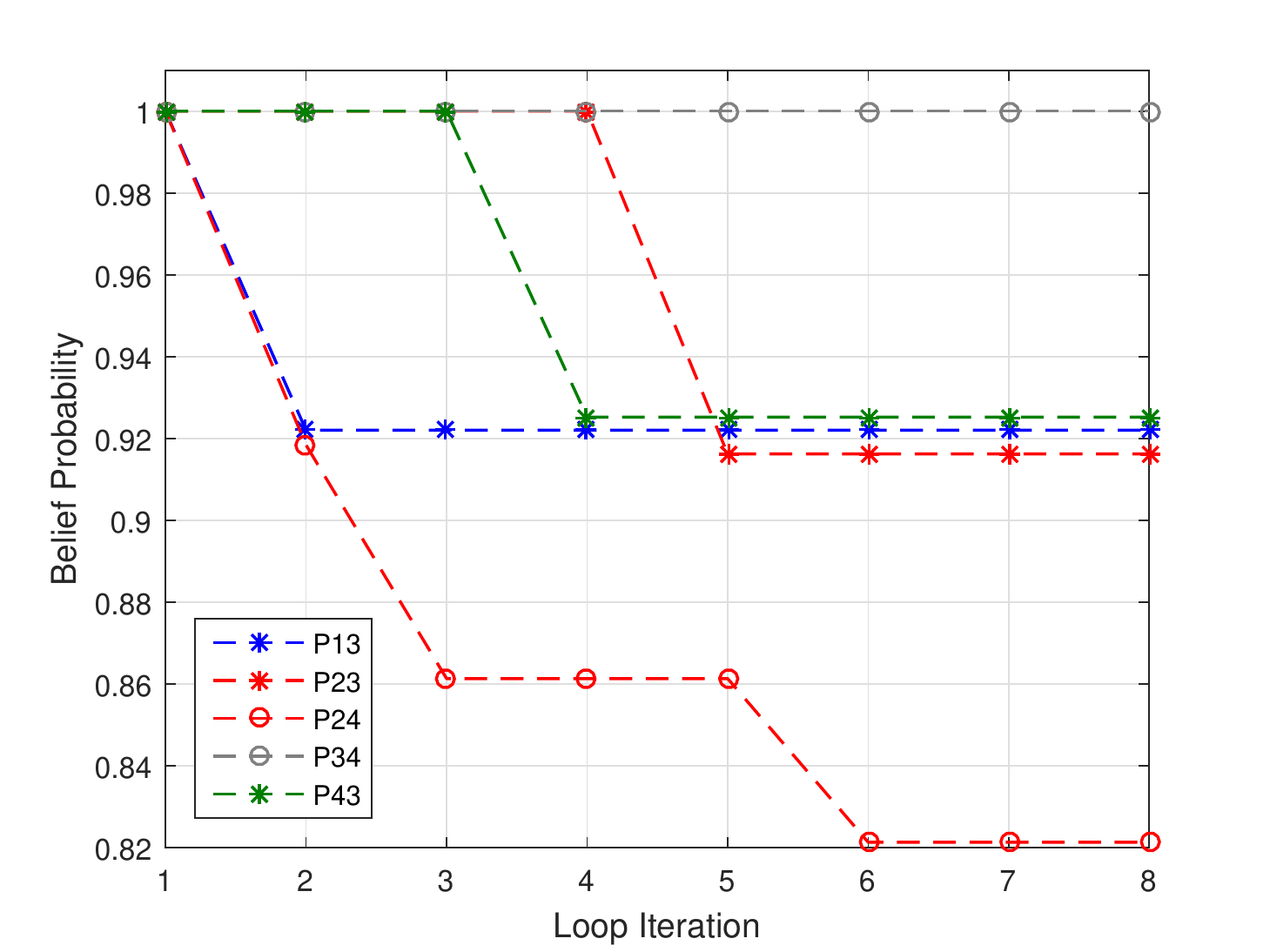}
\caption{Belief probabilities}
\label{fig_update_pro}
\end{subfigure}
\begin{subfigure}[t]{0.5\textwidth}
\centering
\includegraphics[width=\textwidth]{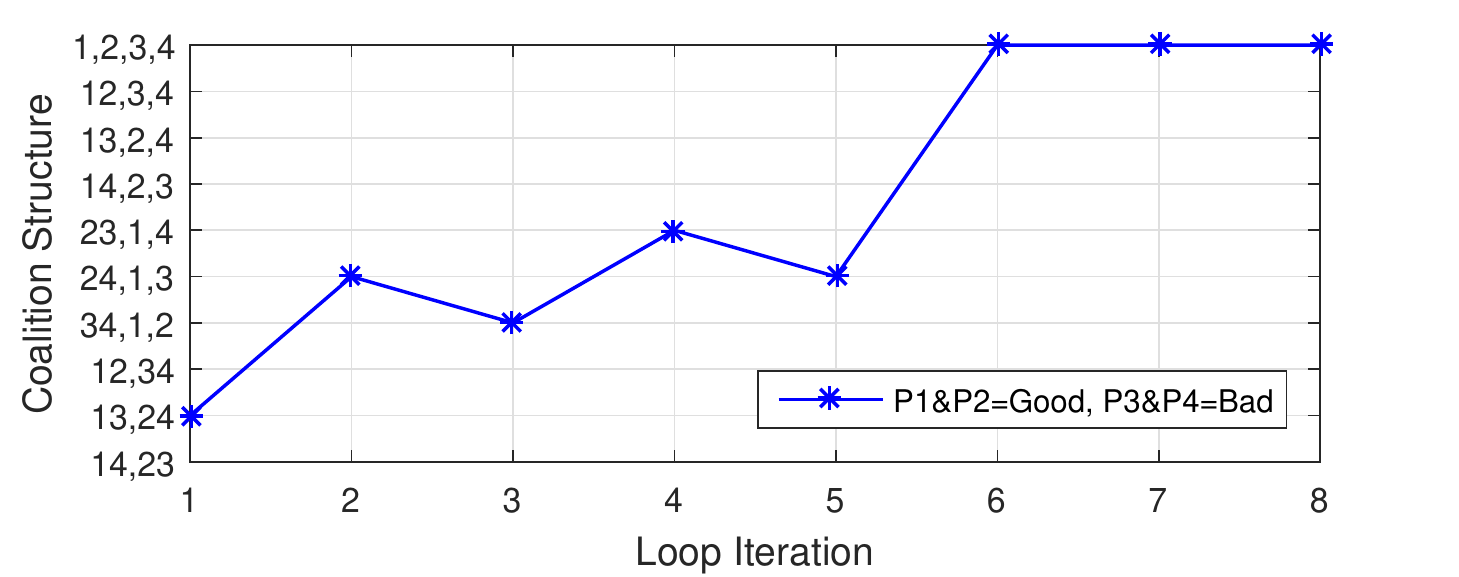}
\caption{Stable coalition formation}
\label{fig_update_coalition}
\end{subfigure}
\caption{Belief update mechanism, where shippers $p_1$ and $p_2$ always deliver packages, shippers $p_3$ and $p_4$ deliver only 75\% of packages that belong to the other cooperative shippers.}
\label{fig_update_main}
\end{figure}

Figure~\ref{fig_update_main} shows the belief probabilities and the stable coalition structure over iterations, i.e., until the belief probabilities converge. From Figure~\ref{fig_update_coalition}, at the first iteration, shippers $p_1$ and $p_3$ cooperate with each other, and shippers $p_2$ and $p_4$ cooperate with each other, i.e., $\Phi_{9}$. Since shipper $p_1$ transfers $11$ packages to shipper $p_3$ and shipper $p_3$ only delivers $8$ of them, the belief probability of shipper $p_1$ to shipper $p_3$ decreases, i.e., $P(b_{1,3} = {\mathrm{T^G}}) = 0.92$, at the second iteration. Note that shipper $p_3$ does not transfer any package to shipper $p_1$ at the first iteration. Similarly, the belief probability of shipper $p_2$ to shipper $p_4$ also decreases, i.e., $P(b_{2,4} = {\mathrm{T^G}}) = 0.92$ at the second iteration. Thus, at the second iteration, the stable coalition structure changes to $\Phi_{6}$. Shipper $p_1$ does not want to cooperate with shipper $p_3$ anymore, but shipper $p_2$ still cooperates with shipper $p_4$. The stable coalition structure is also updated according to the belief probabilities. The belief probabilities are continuously updated based on the observation of the unsuccessful delivery. After the sixth iteration, the belief probabilities converge, and all the shippers have learned about the other shippers' behavior. In this case, all the shippers do not want to cooperate with each other, i.e, coalition structure $\Phi_1$ is reached.

\section{Conclusion and Future works}
\label{sec_conclusion}
We have proposed the Bayesian Shipper Cooperation in Stochastic Drone Delivery (BCoSDD) framework. The uncertainties of drone breakdown and shipper misbehavior have been considered in this framework. The goals of this framework are to help multiple shippers (i) plan their package delivery by either using drones to serve customers or outsourcing the delivery to a carrier and (ii) decide whether the shippers should cooperate or not to minimize the delivery cost. To address the problem, we have proposed (i) package assignment, (ii) cost management, and (iii) shipper cooperation management in the BCoSDD framework. We have formulated the package assignment as the multistage stochastic programming. In the shipper cooperation management, we have used the merge and split algorithm to reach a stable coalition structure. The Bayesian game approach has been utilized to address the uncertainty of shippers' misbehavior. For a shipper that does not have prior knowledge about the other cooperative shippers, we have proposed the belief update mechanism. The belief probabilities will be updated according to the observation when each delivery is performed. At the end, the extensive experiments of the BCoSDD framework have been conducted by using two datasets, i.e., Solomon Benchmark suite and a real dataset from a Singapore logistics industry. Based on the experiments, we can conclude that the BCoSDD framework can achieve the most effective solution, e.g., the lowest cost, compared with other baseline approaches.


\section{Acknowledgment}
 This work was partially supported by Singapore Institute of Manufacturing Technology-Nanyang Technological University (SIMTech-NTU) Joint Laboratory and Collaborative research Programme on Complex Systems.

\begin{IEEEbiography}
 [{\includegraphics[width=1in,height=1.25in,clip,keepaspectratio]{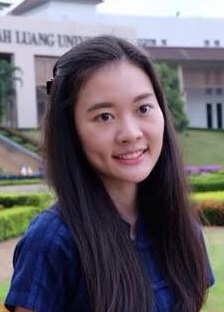}}]{Suttinee Sawadsitang} received her B.Eng in Computer Engineering from King Mongkuts University of Technology Thonburi (KMUTT), Thailand in 2012 and M.Eng from Shanghai Jiao Tong University, China in 2015. She is currently pursuing a Ph.D. degree at SIMTech-NTU Joint Lab on Complex
Systems, Nanyang Technological University, Singapore. Her research interests are in the area of transportation systems, operations research, optimization, and high performance computing.

\end{IEEEbiography}

\begin{IEEEbiography}
 [{\includegraphics[width=1in,height=1.25in,clip,keepaspectratio]{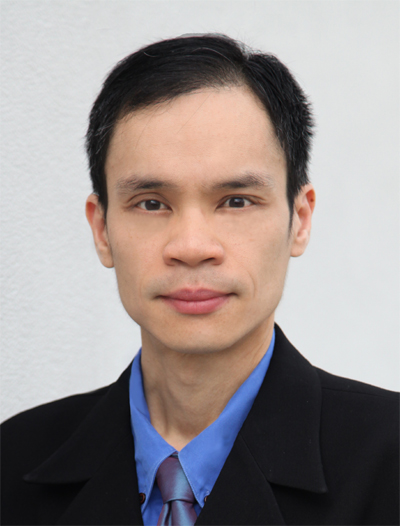}}]{Dusit Niyato}
(M'09-SM'15-F'17) is currently a professor in the School of Computer Science and Engineering, at Nanyang Technological University, Singapore. He received B.Eng. from King Mongkuts Institute of Technology Ladkrabang (KMITL), Thailand in 1999 and Ph.D. in Electrical and Computer Engineering from the University of Manitoba, Canada in 2008. His research interests are in the area of energy harvesting for wireless communication, Internet of Things (IoT) and sensor networks.
\end{IEEEbiography}

\begin{IEEEbiography}
 [{\includegraphics[width=1in,height=1.25in,clip,keepaspectratio]{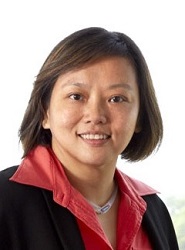}}]
{Puay Siew Tan} received the Ph.D. degree in computer science from the School of Computer Engineering, Nanyang Technological University, Singapore. She is presently an Adjunct Associate Professor with the School of Computer Science and Engineering, Nanyang Technological University and also the Co-Director of the SIMTECH-NTU Joint Laboratory on Complex Systems. In her full-time job at Singapore Institute of Manufacturing Technology (SIMTech), she leads the Manufacturing Control Tower\textsuperscript{TM} (MCT\textsuperscript{TM}) as the Programme Manager. She is also the Deputy Division Director of the Manufacturing System Division. Her research interests are in the cross-field disciplines of Computer Science and Operations Research for virtual enterprise collaboration, in particular sustainable complex manufacturing and supply chain operations in the era of Industry 4.0.
\end{IEEEbiography}

\begin{IEEEbiography}
 [{\includegraphics[width=1in,height=1.25in,clip,keepaspectratio]{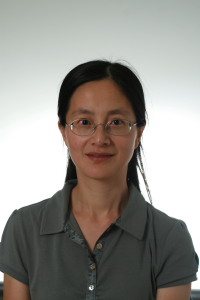}}]
{Ping Wang} (M’08, SM’15) received the PhD degree in electrical engineering from University of Waterloo, Canada, in 2008. She was with Nanyang Technological University, Singapore. Currently she is an associate professor at the department of Electrical Engineering and Computer Science, York University, Canada. Her current research interests include resource allocation in multimedia wireless networks, cloud computing, and smart grid. She was a corecipient of the Best Paper Award from IEEE Wireless Communications and Networking Conference (WCNC) 2012 and IEEE International Conference on Communications (ICC) 2007. She served as an Editor of IEEE Transactions on Wireless Communications, EURASIP Journal on Wireless Communications and Networking, and International Journal of Ultra Wideband Communications and Systems.
\end{IEEEbiography}

\begin{IEEEbiography}
 [{\includegraphics[width=1in,height=1.25in,clip,keepaspectratio]{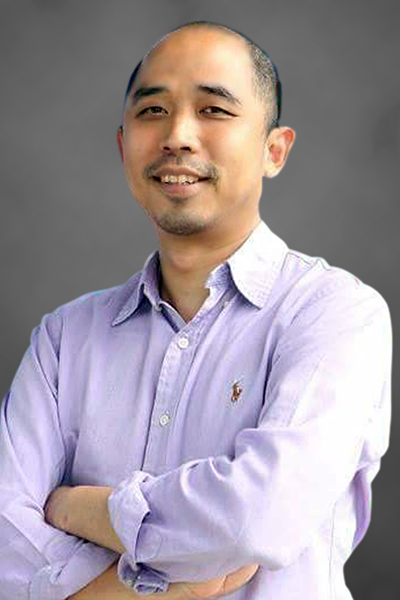}}]
{Sarana Nutanong} received the Ph.D. degree in computer science from the University of Melbourne, Australia, in 2010. He is currently an Assistant Professor with the Department of Computer Science, City University of Hong Kong (CityU), Kowloon, Hong Kong. Before joining CityU, he was a Postdoctoral Research Associate at the University of Maryland Institute for Advanced Computer Studies between 2010 and 2012 and held a research faculty position with the Johns Hopkins University from 2012 to 2013. His research interests include scientific data management, data-intensive computing, spatial–temporal query processing, and large-scale machine learning.
\end{IEEEbiography}

\end{document}